\documentclass[acmsmall,screen]{acmart}
\AtBeginDocument{%
  \providecommand\BibTeX{{%
    \normalfont B\kern-0.5em{\scshape i\kern-0.25em b}\kern-0.8em\TeX}}}
\setcopyright{acmcopyright}
\copyrightyear{2022}
\acmYear{2022}

\acmJournal{CSUR}

\renewcommand\footnotetextcopyrightpermission[1]{} 
\setcopyright{none}

\newcommand{\tabincell}[2]{\begin{tabular}{@{}#1@{}}#2\end{tabular}}  %表格自动换行

\usepackage{subfigure}
\usepackage{supertabular}
\usepackage{tabu}
\usepackage{booktabs}
\usepackage{tabularx}

\usepackage{graphicx}
\usepackage{multirow}

\usepackage{makecell}

\newenvironment{packed_itemize}{
\begin{list}{\labelitemi}{\leftmargin=1.5em}
  \setlength{\itemsep}{1pt}
  \setlength{\parskip}{0pt}
  \setlength{\parsep}{0pt}
  \setlength{\headsep}{0pt}
  \setlength{\topskip}{0pt}
  \setlength{\topmargin}{0pt}
  \setlength{\topsep}{0pt}
  \setlength{\partopsep}{0pt}
}{\end{list}}

\begin{document}

\title{An Interdisciplinary Survey on Origin-destination Flows Modeling: Theory and Techniques}

\author{Can~Rong}
\email{rc20@mails.tsinghua.edu.cn}
\author{Jingtao~Ding}
\email{dingjt15@tsinghua.org.cn}
\author{Yong~Li}
\email{liyong07@tsinghua.edu.cn}
\affiliation{%
  \institution{Beijing National Research Center for Information Science and Technology (BNRist), Department of Electronic Engineering, Tsinghua University}
  \country{China}
  }

\begin{abstract}
Origin-destination~(OD) flow modeling is an extensively researched subject across multiple disciplines, such as the investigation of travel demand in transportation and spatial interaction modeling in geography. 
However, researchers from different fields tend to employ their own unique research paradigms and lack interdisciplinary communication, preventing the cross-fertilization of knowledge and the development of novel solutions to challenges. 
This article presents a systematic interdisciplinary survey that comprehensively and holistically scrutinizes OD flows from utilizing fundamental theory to studying the mechanism of population mobility and solving practical problems with engineering techniques, such as computational models. 
Specifically, regional economics, urban geography, and sociophysics are adept at employing theoretical research methods to explore the underlying mechanisms of OD flows. They have developed three influential theoretical models: the gravity model, the intervening opportunities model, and the radiation model. These models specifically focus on examining the fundamental influences of distance, opportunities, and population on OD flows, respectively. In the meantime, fields such as transportation, urban planning, and computer science primarily focus on addressing four practical problems: OD prediction, OD construction, OD estimation, and OD forecasting. Advanced computational models, such as deep learning models, have gradually been introduced to address these problems more effectively. We have constructed the benchmarks for these four problems at https://github.com/tsinghua-fib-lab/OD\_benckmark. 
Finally, based on the existing research, this survey summarizes current challenges and outlines future directions for this topic. 
Through this survey, we aim to break down the barriers between disciplines in OD flow-related research, fostering interdisciplinary perspectives and modes of thinking.
\end{abstract}


\maketitle

\section{Introduction}

%A city can be characterized as a densely populated and multifaceted environment that fosters the convergence of numerous resources, resulting in the development of shared infrastructure and services~\cite{batty2008size,bettencourt2007growth,jacobs2016death,glaeser2012triumph}. This enables individuals to effectively utilize the available space for diverse activities and collaborations, ultimately leading to optimized production and living standards~\cite{gehl2013cities,bettencourt2007growth,jacobs2016death,whyte1980social,glaeser2012triumph}. 
For land utilization optimization and efficient collaboration, cities have evolved with the delineation of different functional zones, which cater to the diverse living and working demands of citizens. This drives significant interplay between individuals and urban spaces, as they traverse to different functional zones within the city in order to fulfill a myriad of activities that satisfy their unique needs~\cite{hillier1989social,montgomery1998making}. 
%On the other hand, the various functions and services in urban areas are fundamentally sustained by humans, who are the mainstay of the city's operational machinery through traveling to specific workplaces and contributing their labor. Thus, population mobility constitutes a fundamental basis for enabling the smooth functioning of urban ecosystems~\cite{bettencourt2007growth,gonccalves2011modeling,song2010limits}. 
Consequently, the spatial distribution of multiple variables (such as population, vehicular traffic, infectious diseases, etc.) undergoes significant transformations due to population mobility, giving rise to critical challenges such as traffic congestion and epidemic propagation~\cite{banister2011quantification,balcan2009multiscale,storper2006behaviour}. Accordingly, achieving a thorough comprehension of the intricate mechanisms of population mobility within urban environments, and establishing robust models to accurately articulate such complexities, holds great significance for an array of crucial urban applications within the realm of research and applied practice~\cite{song2010modelling,gonzalez2008understanding,batty2007cities,zheng2013u}.

OD~(origin-destination) flows serve as a fundamental manifestation of population mobility. Employing regions or locations as the basic units, they depict the movement of individuals between regions within urban environments~\cite{iqbal2014development,wang2019origin,van1980most,liu2020learning}. 
%In addition to OD flows, other research focuses on inflows and outflows of people in urban areas~\cite{lin2019deepstn+,zhang2017deep,guo2019attention} or broader metrics such as total urban travel volume and total travel distance~\cite{cervero1997travel}. These aspects of population mobility can be further aggregated and calculated from OD flows. Therefore, OD flows that encompass a greater breadth of detailed and nuanced information, rendering it more salient for practical applications across a wider range of contexts, offer greater research value compared to these other kinds of population mobility information. 
This survey centers on the studies on the OD flows, which involves the theoretical research~\cite{haynes2020gravity,simini2012universal,stouffer1940intervening} and practical problem-solving techniques~\cite{liu2020learning,iqbal2014development,wu2018hierarchical,wang2019origin}.

OD flows have been widely studied and demonstrated significant importance in many fields, such as transportation engineering~\cite{cats2014dynamic,lattman2016development,zhan2013urban}, urban planning~\cite{helminen2012commuting,yang2020understanding,credit2022method,malone2001planning}, public health~\cite{balcan2009multiscale,jia2020population,huang2020rapid,kraemer2020effect}, urban geography~\cite{haynes2020gravity,roy2003spatial,wilson1971family}, regional economics~\cite{lesage2009spatial} and social physics~\cite{simini2012universal}. The interplay between OD flow and urban simulation has gained recognition~\cite{zhang2022mirage}. By incorporating OD flow data into computational models, researchers can gain insights into the complex interactions between transportation, infrastructure supply and demand, as well as urban development. Urban simulation represents a new paradigm for studying urban science in the computational era, and the population mobility captured by OD flows plays a crucial role within it.

% \begin{figure*}
%     \centering
%     \includegraphics[width=0.98\textwidth]{figure/ODflows.png}
%     \caption{A diagram of origin-destination flows for comparison with other kinds of mobility data.}
%     \label{intro:OD}
% \end{figure*}

\subsection{The Conception of Origin-destination Flows}

OD flows, which refer to movements of individuals between every two regions in the city at population-level~\cite{simini2021deep,barbosa2018human,luca2021survey}. Specifically, the whole space of the city is usually split into non-overlapping regions $\mathcal{R} = \{ r_i | i=1,...,N \}$. OD flows contain the directed mobility information comprising origins and destinations, and the corresponding numbers of traveling individuals, which could be formulated as follow,
\begin{equation*}
    \{  f_{ij} | r_i \; and \; r_j \in \mathcal{R} \},
\end{equation*}
where~$f_{ij}$ means the volume of flow from region~$r_i$ to region~$r_j$,~$i$ and~$j$ are the indicators of origin and destination respectively. Typically, the OD flows are organized in the form of an OD matrix~$\mathbf{F}$ shown below,
\begin{equation*} \label{eq:ODmatrix}
    \mathbf{F} = 
    \begin{bmatrix} 
        f_{11} & f_{12} & ...    & f_{1N} \\ 
        f_{21} & f_{22} & ...    & f_{2N} \\
        \vdots & \vdots & \ddots & \vdots \\
        f_{N1} & f_{N2} & ...    & f_{NN}
    \end{bmatrix},
\end{equation*}
in which each element represents the flow~$f_{ij}$ between a specific pair of regions. An OD matrix generally represents the mobility flow between all regions within an entire city.

% \begin{figure*}
%     \centering
%     \includegraphics[width=0.65\textwidth]{figure/ODCat.png}
%     \caption{A taxonomy of origin-destination flows.}
%     \label{intro:ODcat}
% \end{figure*}

OD flows can be categorized based on various perspectives, such as temporal nature~\cite{cascetta1993dynamic,liu2020learning,wang2019origin}, the mode of transportation~\cite{bachir2019inferring,gonzalez2020detailed,jiang2022deep,zhang2021short} and types of activities involved~\cite{gonccalves2011modeling,alexander2015origin,yang2015origin}. Based on their temporal characteristics, OD flows can be classified into static and dynamic. The former refers to flows that exhibit a stable structure over a certain period, such as commuting OD flows~\cite{liu2020learning,sadeghinasr2019estimating,liu2020urban,yin2023convgcn}, while the latter captures the time-varying patterns of population mobility, which reflects the dynamic changes in the travel behavior of individuals~\cite{bachir2019inferring,cao2021day,wang2019origin}. OD flows can be classified based on modes of travel, including bus OD flows~\cite{gonzalez2020detailed}, railway OD flows~\cite{jiang2022deep}, OD flows of vehicles on the road networks~\cite{bera2011estimation}, etc. 
%These subcategories reflect the different types of transportation modes that individuals may use when traveling from one region to another and have implications for the analysis and modeling of travel behavior. 
Various studies~\cite{gonccalves2011modeling,alexander2015origin,yang2015origin} have categorized OD flows based on different types of activities, resulting in two primary categories: commuting OD flows~\cite{liu2020learning,sadeghinasr2019estimating,liu2020urban,yin2023convgcn}, which relate to travel for work purposes, and non-commuting OD flows~\cite{yang2015origin}, which encompass all other types of travel, such as leisure, social, and shopping activities. This distinction offers insights into the patterns and behavior of travel demand and highlights the importance of accounting for variability in travel purposes when analyzing OD flows.

\subsection{The Research on OD Flows Across Disciplines}

In light of the significant importance of OD flows it is worth noting that their traditional methods of acquisition, such as roadside and household surveys, tend to be highly time-consuming and resource-intensive~\cite{axhausen2002observing,schuessler2009processing,iqbal2014development}, which have driven the relevant academic disciplines and fields to employ their research paradigms to investigate the mechanism and modeling of human mobility behind OD flows~\cite{simini2021deep,simini2012universal,iqbal2014development,van1980most,wang2019origin,liu2020learning}. The current research on OD flows can be categorized into two main aspects. The first aspect involves studying the underlying mechanisms of population mobility from a theoretical perspective. In contrast, the second aspect focuses on addressing practical challenges related to OD flows. Detailed theoretical research schema of these fields are introduced as flows.

% The research of OD flows through the lens of theoretical paradigms and holds relevance across the fields of urban geography, regional economics, and sociophysics. Each discipline applies distinct theoretical frameworks to elucidate the fundamental mechanisms driving population interactions between regions, i.e., OD flows. We will delineate the association between OD flows and the separate academic disciplines alongside their respective motivation for research and further elaboration on this specific research topics will be provided in Sec \ref{sec:theory}.

\begin{packed_itemize}
    \item \textbf{Urban Geography and Regional Economics.} Urban geography and regional economics employ theoretical research methodologies to investigate the OD flows from a spatial perspective~\cite{anas1998urban,alumni2013economics,krugman1991increasing,batty2008size}. One crucial research involves spatial interaction modeling which studies the interaction among diverse regions in urban contexts such as the flows of people, trade, communication, etc~\cite{haynes2020gravity,wilson1971family,roy2003spatial,wegener2004land,lesage2009spatial}. Flows of individuals in the shape of OD flows are regarded as a critical object of research within this domain.
    \item \textbf{Social Physics.} Scholars specializing in urban science within the discipline of sociophysics~\cite{barthelemy2011spatial,gonzalez2008understanding,louf2014congestion,simini2012universal} frequently employ physical models or adopt the research paradigm of physics to investigate population mobility in cities, aiming to provide a profound elucidation of the underlying physical mechanisms that give rise to the observed mobility patterns.
\end{packed_itemize}

\begin{figure*}
    \centering
    \includegraphics[width=0.98\textwidth]{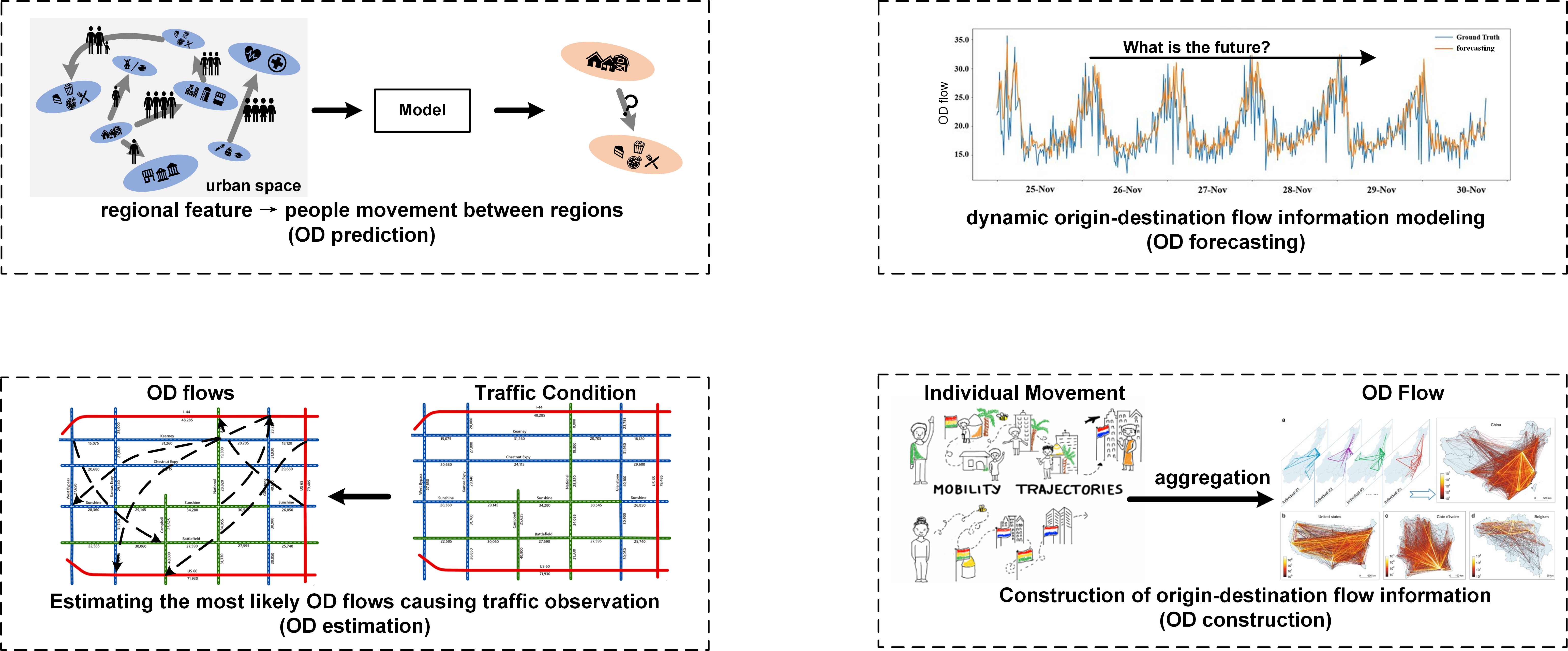}
    \caption{Research on origin-destination flows across different domains. The figure illustrates four classical practical issues related to OD flows, each of which possesses strong interdisciplinary attributes. To explicitly showcase the contribution of each discipline to these problems, we employ a radar chart to interpret the contribution degree of each discipline to the respective issue.}
    \label{intro:ODfields}
\end{figure*}

The study of OD flows also encompasses a range of practical issues, particularly those arising from the fields of transportation and urban planning, such as the problems shown in Fig. \ref{intro:ODfields}. Advancements in computer science, electronic communication, statistics, and systems engineering have greatly contributed to the development and application of advanced technologies to address these challenges, yielding promising outcomes in solving techniques and application scenarios~\cite{liu2020learning,iqbal2014development,yin2023convgcn,wang2019origin}. The following presents a comprehensive explication of the relationships existing between these domains and OD flows.
\begin{packed_itemize}
    \item \textbf{Computer Science.} The utilization of data mining methodologies to tackle challenges in urban contexts represents a burgeoning area within computer science known as urban computing~\cite{zheng2014urban}. An integral facet of this research involves the employment of data-driven models that are capable of accurately characterizing and modeling the OD flows, such as the OD prediction and OD forecasting problems shown in Fig. \ref{intro:ODfields}.
    \item \textbf{Transportation.} The term OD flows was originally proposed by the transportation field as input to numerous transportation problems, and thus, holds a significant position in transportation research, making it the area where OD flows have been extensively studied. The vast majority of human travel is reliant upon transportation systems, thus leading to the explicit depiction of travel patterns within such systems. In this context, the transportation system serves as an observatory that captures extensive traces of human travel. Transportation researchers have explored these traces to estimate the OD flows~\cite{van1980most,munizaga2012estimation,cascetta1993dynamic,wu2018hierarchical}, as shown in the OD estimation problem in Fig.\ref{intro:ODfields}.
    \item \textbf{Urban Planning.} The relationship between urban planning layout and population mobility is highly dependent on one another~\cite{cervero1997travel,ewing2010travel,crane2000influence,handy2002built}. In order to make informed adjustments, planners must conduct a thorough analysis of OD flows to gain insight into the strengths and shortcomings of current urban plans~\cite{cervero1997paradigm}. The main problem faced by urban planning is OD construction.
    
\end{packed_itemize}

\begin{figure*}
    \centering
    \includegraphics[width=0.98\textwidth]{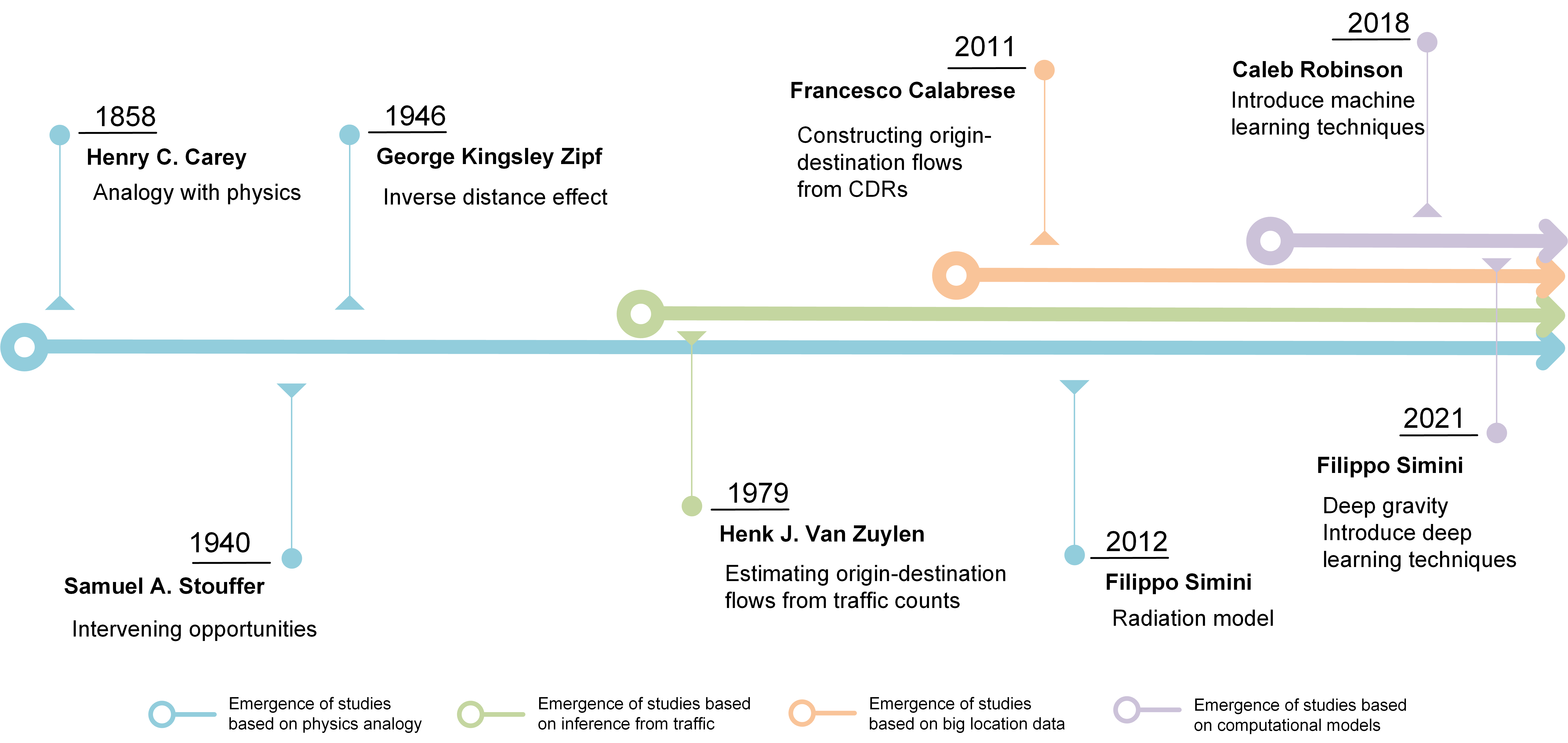}
    \caption{A timeline of important research on origin-destination flows in diverse academic disciplines.}
    \label{intro:timeline}
\end{figure*}

As presented previously, the OD flows have been extensively researched across multiple fields, highlighting the strong interdisciplinary nature of this research. Despite the existence of disparate terms adopted in different disciplines, such as spatial interaction and population flow, this paper solely employs the universally recognized term, namely OD flows, to mitigate any potential ambiguity. Subsequently, we present a comprehensive timeline that chronicles the research history of the OD flows in diverse academic disciplines, as shown in Fig. \ref{intro:timeline}. 
The genesis of the study of OD flows is primarily associated with geography. Henry C. Carey pioneered the investigation of the physics analogy of OD flows~\cite{carey1859principles} in 1858. Subsequently, the celebrated linguist Zipf formalized the gravity model~\cite{zipf1946p} in 1946, which has since gained widespread recognition as a seminal model in research of OD flows. 
% Subsequent research in the 20th century~\cite{erlander1990gravity,stewart1941inverse,huff1963probabilistic,lakshmanan1965retail,harris1978equilibrium,wilson2011entropy,evans1973relationship} endeavored to enhance gravity models by integrating supplementary information. 
With the rapid growth of public transportation and the pressing need to address traffic congestion and environmental pollution~\cite{banister2008sustainable,newman1999sustainability,litman2012evaluating}, researchers in the transportation field began to explore the feasibility of utilizing traffic observations, to estimate OD flows~\cite{van1980most,wu2018hierarchical,cascetta1993dynamic,yang2017origin}. 
% This approach entails identifying the OD flows that are most likely to result in the observed traffic status. 
The proliferation of ICT (Information And Communication Technology) and mobile network communication devices has led to a vast accumulation of data pertaining to human mobility behavior. As a result, this influx of data has facilitated the study of OD flows~\cite{iqbal2014development,calabrese2011estimating,alexander2015origin}. The famous planner Carlo Ratti proposed to leverage Call Detail Records (CDR) data to devise OD matrices~\cite{calabrese2011estimating}, which has sparked a considerable amount of effort~\cite{iqbal2014development,alexander2015origin,toole2015path,pan2006cellular,caceres2007deriving,yang2015origin} from researchers and practitioners in this direction. 
Sociophysics scientists have been able to uncover the mechanism governing human mobility~\cite{song2010modelling,gonzalez2008understanding}, such as the radiation model~\cite{simini2012universal}. 
%In addition to employing an analogy of radiation and absorption processes in solid-state physics to model human mobility~\cite{simini2012universal}.
With the development of machine learning, computer scientists have increasingly investigated the use of sophisticated models to capture complex spatiotemporal correlations between OD flows and urban space~\cite{liu2020learning,yin2023convgcn,yao2020spatial,koca2021origin,cai2022spatial}.
% This approach has yielded markedly superior results in terms of predictive accuracy. Tree-based machine learning technique~\cite{friedman2001greedy,breiman2001random} offers robustness and highly capable high-dimensional nonlinear fitting ability that leads to significant improvement in modeling OD flows on urban space as compared to conventional models~\cite{pourebrahim2019trip,robinson2018machine}. The incorporation of graph neural networks~\cite{liu2020learning,cai2022spatial,rong2023goddag} and diffusion models~\cite{ho2020denoising,vignac2022digress,} has resulted in a further improvement in the precision of these kinds of models.

\subsection{Related Works}

This part aims to introduce the common parts and dissimilarities between this review and other academic investigations that involve OD flows. These surveys can be categorized into three distinct groups: (1) human mobility, (2) spatial interaction modeling, and (3) OD flow estimation based on observable transportation status. Table \ref{Tab:existingsurvey} has already provided a comprehensive overview of the specific aspects addressed in each review works. The interrelation and distinctions between these aspects and our work are shown in appendix \ref{apdx:related}.

\subsection{Organization}
The initial section of this survey expounds on the fundamental information of OD flows. 
The outline of this survey encompasses data, theory, techniques, and applications, and any pertinent data or information related to these aspects are first summarized in Section \ref{sec:data}.
In particular, the theoretical mechanisms related to OD flows will be systematically presented in Section \ref{sec:theory}. 
Next, the relevant techniques will be classified based on specific practical problems and elaborated upon in Section \ref{sec:tec}.
In Section \ref{sec:app}, specific applications pertaining to OD flows in diverse fields will be explicated.
We undertake an evaluation of the existing challenges and future direction in the study of OD flows in Section \ref{sec:future}.
Finally, we will conclude this survey in Section \ref{sec:conclusion}.

\section{Data Source Requirements} \label{sec:data}

% Due to the interdisciplinary nature of research on OD flows, multiple fields, and disciplines are involved, each utilizing distinct information in their studies. Consequently, the scope of data used in OD flow research is incredibly broad and diverse. 
The data utilized in OD flow research can be broadly classified into two categories. The first category comprises  mobility data related to OD flows. The second category encompasses auxiliary data, which is leveraged to explore the association between OD flows and other urban spatiotemporal characteristics.

\subsection{Mobility Data Related to OD Flows}

\begin{table}[]
\resizebox{9cm}{!}{
\begin{tabular}{@{}ccccc@{}}
\toprule
                     & CDRs & Cellular & GPS          & Location-based Check-ins \\ \midrule
Spatial Granularity  & low  & low      & high         & low                      \\
Temporal Granularity & low  & high     & high         & low                      \\
Population Coverage  & high & high     & low          & high                     \\ \bottomrule
\end{tabular}
}
\caption{The systematic comparison of individual trajectory-level location recording data.}
\label{tab:locdata}
\end{table}

The sources of mobility data are diverse and have undergone continuous evolution with the advancement of technology over time~\cite{simini2021deep,iqbal2014development,calabrese2011estimating,liu2020learning,wang2019origin}, involving survey data, individual trajectories, and transportation records. The individual trajectory data involve Call Detail Records, Cellular Network Access, GPS Records, and Location-based Social Network Check-ins. The transportation records include traffic surveillance video, smart cards, and taxi orders. Detailed introductions to these data sources can be found in the appendix \ref{apdx:data}.
% The conventional method for obtaining OD flow-related information involves household and roadside surveys, which are associated with high time and monetary costs and often yield data with suboptimal temporal validity~\cite{axhausen2002observing,schuessler2009processing}. with the rapid development of ICT, a large number of location sensors were incorporated into mobile devices such as smartphones, and became widely available to urban residents, recording a wealth of individual location information~\cite{iqbal2014development,bera2011estimation,schuessler2009processing}. Through statistical analysis of the trajectories of a significant number of individuals, it is possible to extract meaningful information regarding the patterns of OD flows exhibited by the population. The relevant data involved is as follows:

\begin{figure*}[t]
    \centering
    \subfigure[Individual trajectories.]{
    \label{fig:dataa}
    \includegraphics[width=0.23\textwidth]{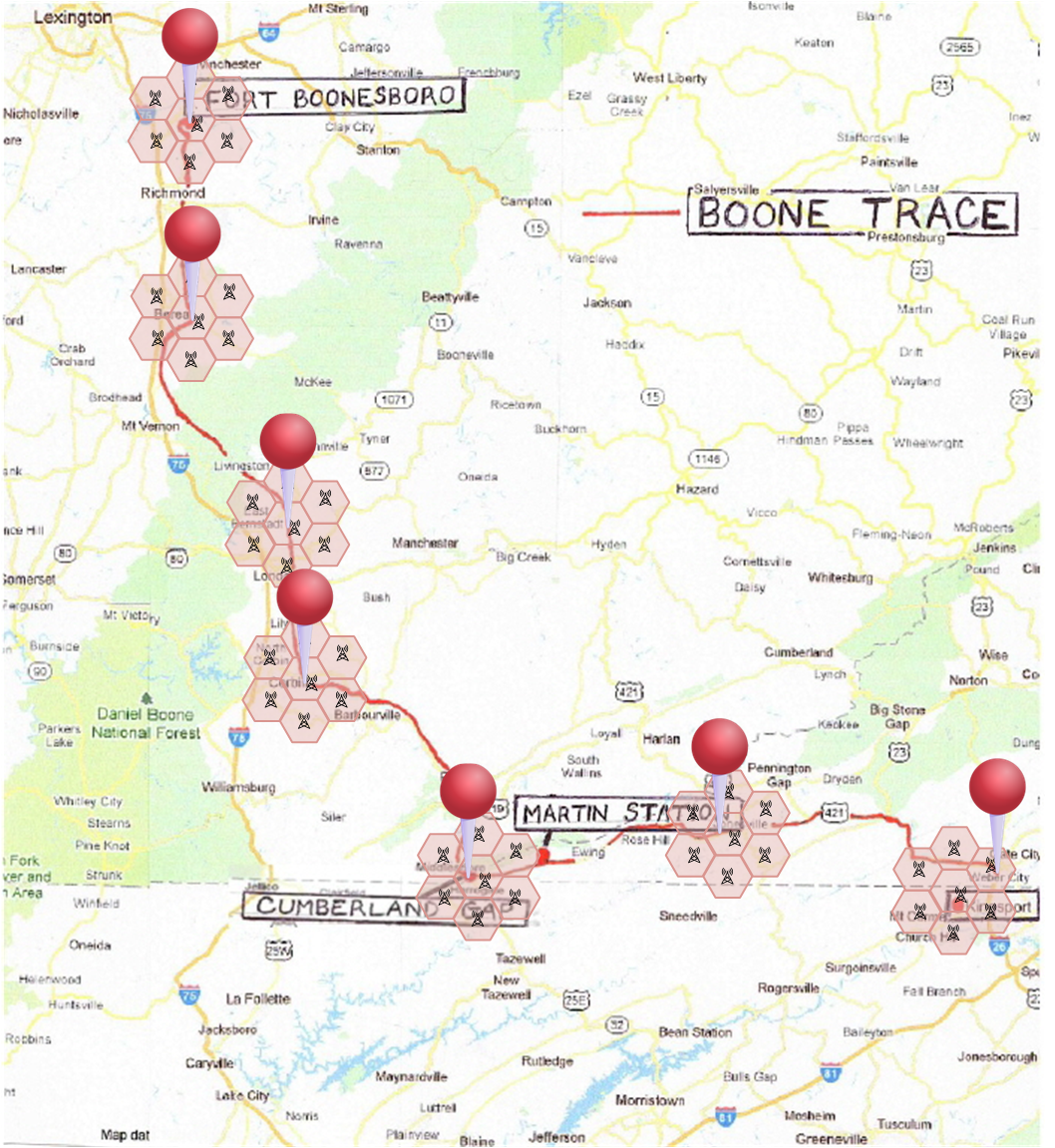}
    }\hfill
    \subfigure[Traffic conditions.]{
    \label{fig:datab}
    \includegraphics[width=0.23\textwidth]{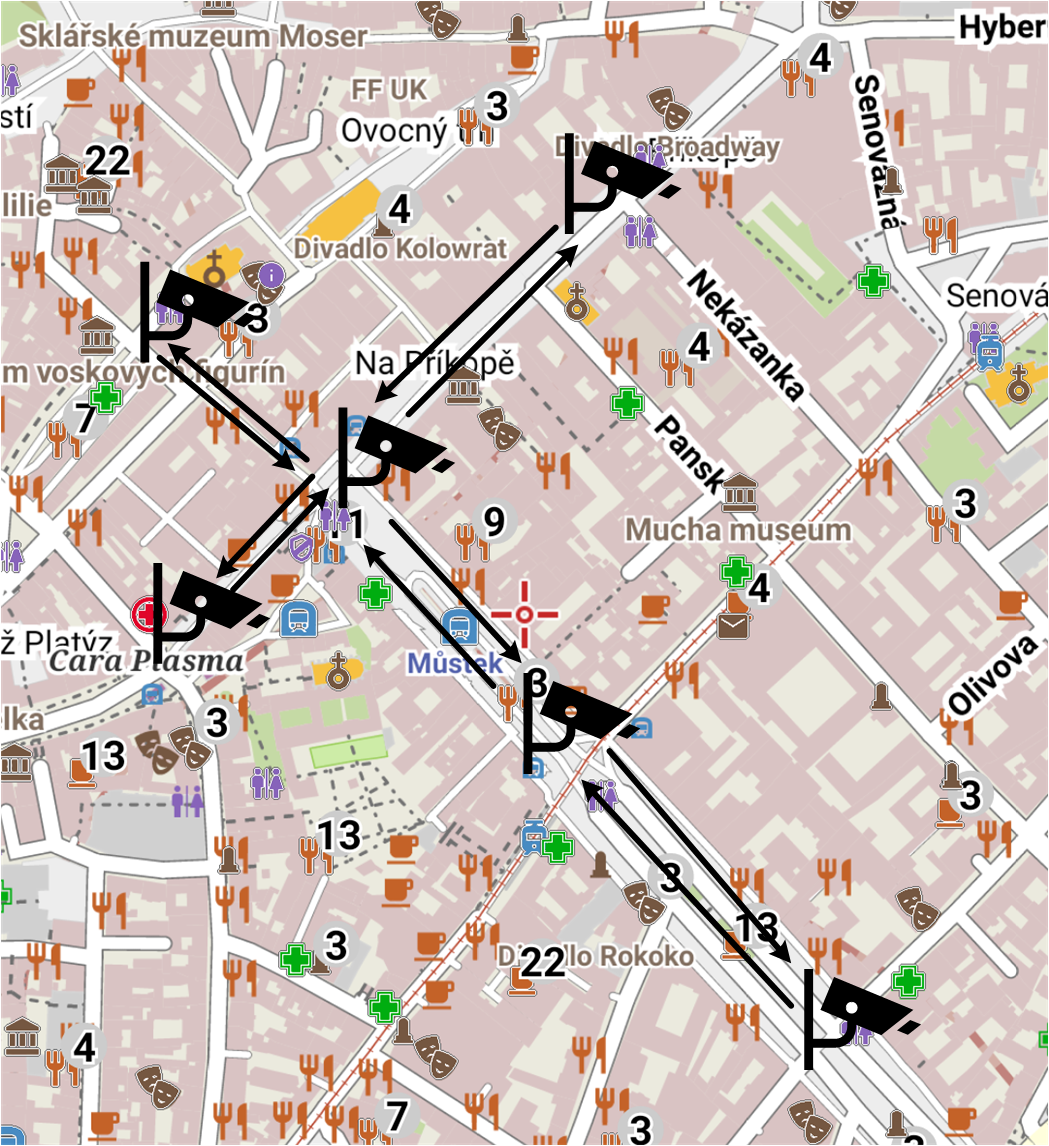}
    }\hfill
    \subfigure[Urban characteristics.]{
    \label{fig:datac}
    \includegraphics[width=0.28\textwidth]{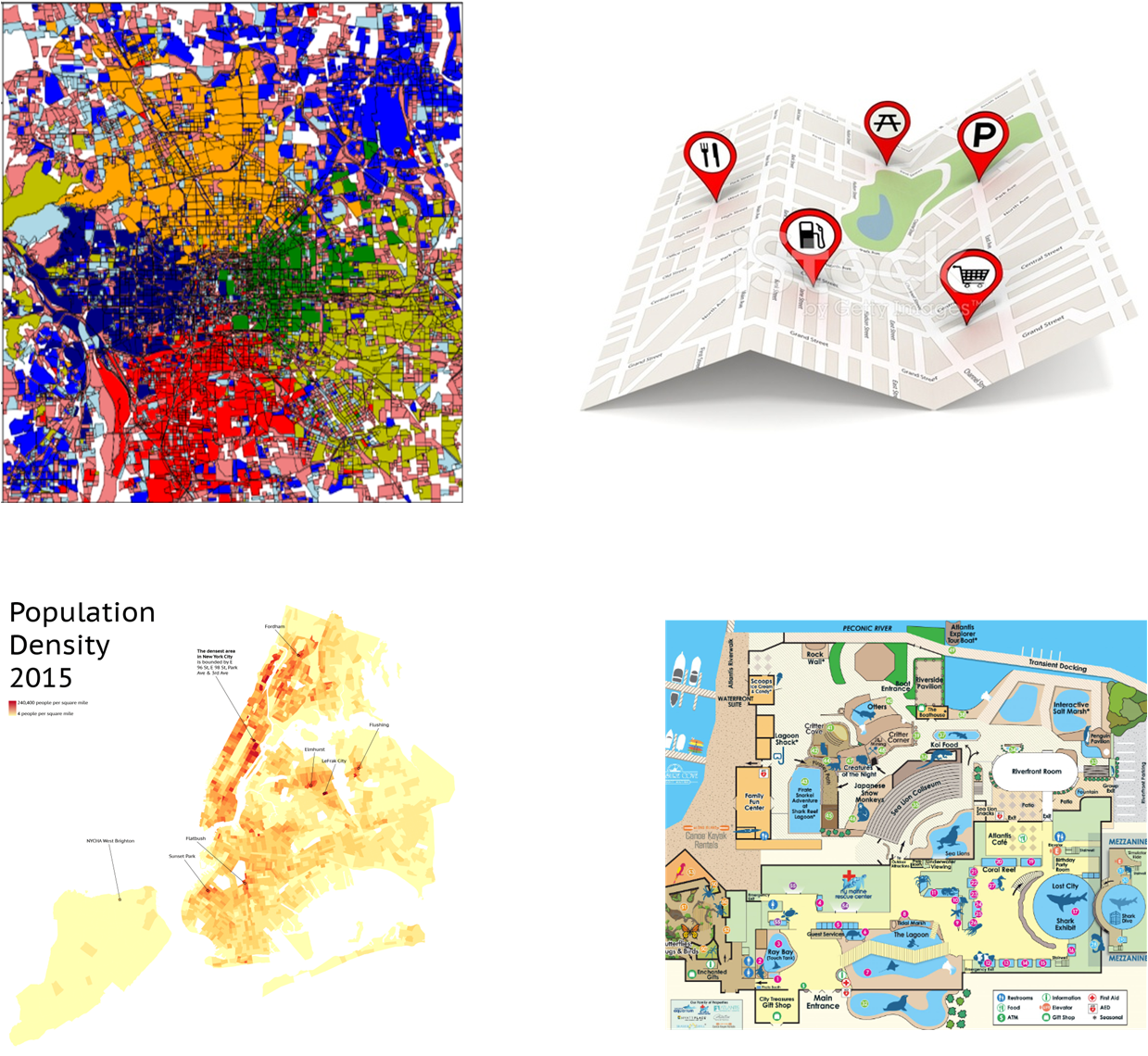}
    }
    \caption{A schematic diagram of commonly used mobility data related to OD flows.}
    \label{data:data}
\end{figure*}

In order to facilitate the comparative analysis of individual trajectories, the pertinent features has been collated and presented in Table \ref{tab:locdata}. It is apparent that the existence of flawless data is elusive. Therefore, when conducting research, it is crucial to meticulously consider and evaluate the specific scenario and its requirements in order to determine the most appropriate and effective data to employ. 
Among them, CDRs data, cellular network access data, and social network check-ins have coarse spatiotemporal granularity and thus are commonly utilized to estimate long-term OD flows, such as commuting OD flows. While GPS data with finer spatiotemporal granularity can enable the investigation of individuals' more detailed mobility patterns, including specific activities, it is high cost and low population coverage renders it unsuitable for conducting citywide research~\cite{kitchin2014real}. This kind of trajectory data is shown in Fig.\ref{fig:dataa}. The vehicle-related OD flows information obtained from sensors and video data of traffic surveillance, as shown in Fig.\ref{fig:datab}, is typically utilized in real-time and dynamic scenarios, such as traffic congestion management and traffic accident prevention~\cite{tian2014hierarchical}. The OD flows obtained from smart card data and taxi order data typically reflect the demand of individuals for a specific mode of transportation, and are often utilized for transportation demand forecasting~\cite{wang2019origin,shi2020predicting}.

\subsection{Auxiliary Data}
Apart from the mobility data mentioned above, the investigation of specific research questions necessitates the inclusion of additional information such as regional socioeconomics and real-time traffic status, as shown in Fig.\ref{data:data}. By utilizing this information, researchers can gain deeper insights into the underlying mechanisms of population mobility~\cite{zheng2014urban}, or use it as auxiliary signals to address practical problems related to OD flows~\cite{stupar2018socio}. Data sources for obtaining such complementary information can be broadly classified into three categories: region-level socioeconomic data, transportation observation data, and urban geographic data. 

\textbf{Region-level Socioeconomic Data.} The mobility of individuals within urban environments is primarily motivated by the desire to participate in a diverse range of activities located across the cityscape, in order to satisfy distinct demands~\cite{hillier1989social,montgomery1998making}. These activities may encompass occupational pursuits, retail consumption, medical treatment, and other associated functions. As such, the regional attributes of urban spaces, as well as the corresponding roles they play, are intimately interconnected with patterns of population mobility. Numerous types of information can be utilized to extract the socioeconomic characteristics of regions within urban areas. Such data often includes demographics, land use patterns, points of interest (POIs), and infrastructure. These factors are frequently incorporated into extant scholarly literature when profiling spatial functions. A detailed introduction of this information is given in appendix \ref{apdx:data}.

%Beyond the factors previously outlined, various other kinds of special conditions, including policy and public health crises and disasters, can exert significant influences on patterns of population mobility. However, as these events are often unforeseen and emerge as sudden disruptions, they are not within the purview of this paper.

\textbf{Observable Transportation Status or Records.} OD flows are considered to be critical information in the field of transportation, as they serve as a proxy for travel demands. 

Drawing on their specialized knowledge and skills, transportation professionals frequently rely on observations of the transportation system to infer OD flows and related travel patterns. This kind of information generally encompasses traffic flow and real-time speed data captured from road networks. 

% In the four-step~\cite{mcnally2007four} (I. trip generation, II. trip distribution, III. routing, IV. mode choice) approach to transportation modeling, the estimation of traffic flow and speed data is typically achieved through the use of dynamic traffic assignment (DTA) models~\cite{peeta2001foundations} in conjunction with simulation tools, given OD flows. 
%The DTA model is designed to allocate OD flows to specific routes in the network, based on an analysis of the available paths. 
% The estimation of OD flows based on observed traffic conditions represents the inverse problem of dynamic traffic assignment (DTA). Given traffic flow and vehicle speed, these two distinct types of information (although they may be interrelated through macroscopic traffic flow models such as the Lighthill-Whitham-Richards model~\cite{lighthill1955kinematic}), the estimation of OD flows based on these variables typically involves the utilization of different techniques and methodologies. Below, we will offer comprehensive descriptions of them.
\begin{packed_itemize}
    \item \textbf{Traffic Flow.} This metric is often referred to as traffic counts or link counts, depending on the specific application or context. Given that traffic flow, each individual vehicle can be associated with a corresponding trip in the OD flows. 
    \item \textbf{Vehicle Speed.} Vehicle speed refers to the rate at which vehicles travel along a particular roadway segment. Compared to traffic flow, vehicle speed is a more instantaneous measure that provides a more fine-grained understanding of traffic conditions. As such, it can be more suitable for modeling dynamic transportation scenarios and capturing short-term fluctuations in traffic demand and congestion.
\end{packed_itemize}
The emergence of intelligent transportation systems~(ITS) has provided a variety of means to collect traffic information. Traditionally, traffic flow and vehicle speed are obtained by deploying sensors on the road network. 
%These sensors can include inductive loops, cameras, or other devices that measure the number of vehicles passing through a specific point or their speed. 
While these methods have been proven to be effective, recent advancements in wireless communication, sensor technology, and data analytics have opened up new possibilities for collecting traffic data using innovative approaches such as crowdsourcing, floating cars with technologies of Bluetooth, unmanned aerial vehicles (UAVs), and radio frequency identification (RFID). The comparative introduction of various data collection methods is presented in Table \ref{tab:trafdata} in Appendix~\ref{apdx:auxdata}.

\section{Theoretical Models and Analysis for OD Flows} \label{sec:theory}

In an endeavor to acquire a more profound comprehension of the underlying mechanisms of OD flows, a multitude of theoretical perspectives have been advanced~\cite{simini2012universal,gonzalez2008understanding,song2010limits,roy2003spatial,huff1963probabilistic}. %These theories try to employ diverse methodologies in order to establish comprehensive frameworks elucidating the intricacies and mechanism of population mobility and explaining the macro phenomenon, which has been represented in OD flows. 

\subsection{Gravity Model}

The gravity model in population mobility modeling is derived from Isaac Newton's law of universal gravitation. 
% This law adeptly elucidates the gravitational forces exerted between objects and enjoyed an important standing in physics. Specifically, if $d_{ij}$ signify the distance separating objects $i$ and $j$, and $m_i$ and $m_j$ correspond to their individual masses, Newton's theory posits that the gravitational force $F_{ij}$ is given by the formula,
% \begin{equation}
%     F_{ij} = G\frac{m_im_j}{d_{ij}^2},
% \end{equation}
% where $G$ stands for the gravitational constant. 
Pertaining to the investigation of interregional interaction, the genesis of notions akin to the gravity model can be traced back to the mid-19th century~\cite{carey1859principles}. Then, it was Zipf's seminal work in 1941~\cite{zipf1946p} that marked the inaugural formalization of the gravity model, accompanied by a theoretical framework. To be more specific, Zipf postulated that the magnitude of interaction, manifested through population or trade flows between two distinct locations, is directly commensurate with characteristics such as the population size, while concurrently exhibiting an inverse correlation with the distance and provided a specific formula to quantify this relationship,
\begin{equation} \label{eq:zipf}
    T_{ij} \propto \frac{P_iP_j}{d_{ij}},
\end{equation}
where $P_i$ and $P_j$ are the number of population of region $i$ and $j$. And the $d_{ij}$ means the distance. Zipf compared his findings with the principles of Newton's gravitational formula~\cite{zipf1946p}, thereby paving the way for the subsequent emergence and progression of a series of studies of gravity models that have come to underpin the spatial interactions and human mobility in later research~\cite{roy2003spatial,wilson1971family,haynes2020gravity}. It was not until the 1960s that Huff incorporated calibratable parameters of distance into the framework, marking a significant development in the model's evolution. Subsequently, calibratable parameters were also assigned to $P_i$ and $P_j$, resulting in the formation of the commonly encountered gravity model formulation seen in later research,
\begin{equation}\label{eq:genG}
    T_{ij} = \lambda \frac{P_i^{\beta_i}  P_j^{\beta_j}}{d_{ij}^\alpha},
\end{equation}
where $\lambda$, $\alpha$, $\beta_i$ and $\beta_j$ represents the four calibratable parameters. 
%Typically, distance provides a decaying effect, so the value of parameter $\alpha$ is commonly greater than 0. 
The introduction of adjustable parameters in the gravity model is of significant importance, as it allows for the effects of various factors, such as differences in population structures, on population mobility to be taken into account. For example, regions with a younger population tend to have a higher production intensity, resulting in a larger $\beta_i$ when used as the origin zone in the gravity model. 
%In contrast, regions with an aging population tend to have lower population mobility, resulting in a smaller $\beta_j$. 
Despite later debates concerning the power and exponential forms of distance arising~\cite{fotheringham1989spatial}, these deliberations primarily address nuanced aspects. A more comprehensive and generalized expression can be exemplified by the following formula,
\begin{equation}
    T_{ij} = \lambda f_i (\mathbf{P}_i) f_j (\mathbf{P}_j) f_d(d_{ij}),
\end{equation}
where the $f_i$ and $f_j$ means any kind of functions map the feature vectors $\mathbf{P}_i$ and $\mathbf{P}_j$ of origin $i$ and destination $j$, and $f_d$ can be any format of effect from distance. To provide a more specific explanation, $f_i(\mathbf{P}_i)$ can describe the production intensity at the origin, while $f_j(\mathbf{P}_j)$ can represent the attraction intensity at the destination in complex scenarios. When the $\mathbf{P}_i$ and $\mathbf{P}_j$ vectors contain only one element respectively, and the function $f$ adopts the power format, it degenerates into Formula \ref{eq:genG}.

The ensuing investigation and advancement of gravity models can be categorized based on two distinct perspectives. Firstly, numerous adaptations of the gravity model have been proposed to cater to the diverse requirements of varying contexts. Secondly, attempts have been made to extrapolate the gravity model from underlying individual principles, thereby providing a robust theoretical foundation for its application.

\subsubsection{Adaptations of Gravity Model Tailored to Various Scenarios and Contexts}
Adjustments to the gravity model in different scenarios are usually made by incorporating additional constraints based on available information. Given the available information, the problem scenario, and the corresponding gravity model, there are five categories of methods that can be used to incorporate constraints. To facilitate the discussion, they are classified into five types, as shown in Tab \ref{theory:gravities-class}. 
%When there is no known information to guide the modeling process, it is referred to as a non-constrained gravity model, which is the ordinary gravity model mentioned previously. 
%When both the outflows of origins and the inflows of destinations are known, the model used is called the doubly-constrained gravity model, which corresponds to the first quadrant in the figure. Therefore, when only one of them is known, it is also referred to as a singly-constrained gravity model in academic literature~\cite{wilson1971family}.

% \begin{figure*}
%     \centering
%     \includegraphics[width=0.45\textwidth]{figure/gravity-four.png}
%     \caption{Classification of gravity models with different available information as constraints.}
%     \label{theory:gravities-class}
% \end{figure*}

\begin{table}[]
\resizebox{12cm}{!}{
\begin{tabular}{|cc|c|c|c|c|}
\hline
\multicolumn{2}{|c|}{\begin{tabular}[c]{@{}c@{}}Gravity Model of \\ Different Constraints\end{tabular}}                       & Inflow                    & Outflow                   & Total                     & Fields and Applications                                                                            \\ \hline
\multicolumn{1}{|c|}{no constraints}                                                                 & non-constrained        & $\times$                  & $\times$                  & $\times$                  & \begin{tabular}[c]{@{}c@{}}urban geographics, \\ regional economics,\\ urban planning\end{tabular} \\ \hline
\multicolumn{1}{|c|}{\begin{tabular}[c]{@{}c@{}}only a total flow\\ constraints\end{tabular}}        & global-constrained     & $\times$                  & $\times$                  & \checkmark & \begin{tabular}[c]{@{}c@{}}urban planning, \\ scenic spot planning\end{tabular}                    \\ \hline
\multicolumn{1}{|c|}{\multirow{2}{*}{\begin{tabular}[c]{@{}c@{}}singly \\ constraints\end{tabular}}} & attraction-constrained & \checkmark & $\times$                  & \checkmark & \begin{tabular}[c]{@{}c@{}}regional economics, \\ transportation\end{tabular}                      \\ \cline{2-6} 
\multicolumn{1}{|c|}{}                                                                               & production-constrained & $\times$                  & \checkmark & \checkmark & \begin{tabular}[c]{@{}c@{}}computer science, \\ transportation\end{tabular}                        \\ \hline
\multicolumn{1}{|c|}{\begin{tabular}[c]{@{}c@{}}doubly \\ constraints\end{tabular}}                  & doubly-constrained     & \checkmark & \checkmark & \checkmark & transportation                                                                                     \\ \hline
\end{tabular}
}
\caption{A systematic comparison between different kinds of constraints of gravity models. For easy to read, we abbreviate the non-constrained gravity model to non-constrained, and so on with the others.}
\label{theory:gravities-class}
\end{table}

\textbf{Globally-constrained Gravity Model.} The globally-constrained gravity model is an advanced form of the gravity model that incorporates constraints on the total number of trips or movements throughout the entire study area. Below is the symbolic representation,
\begin{equation}
    \sum_{i=1}^m \sum_{j=1}^{n} \hat{T}_{ij} = \sum_{i=1}^m \sum_{j=1}^{n} T_{ij} = T ,
\end{equation}
where $m$ and $n$ denote the number of origins and destinations respectively, and $T$ means the number of total trips. 

In transportation planning, the total amount of travel in a study area can often be obtained. This quantity does not need to be completely accurate. Furthermore, there are instances where more granular settings, such as tourism planning within designated scenic areas and urban commuting patterns, allow for the acquisition of total trip information. 

\textbf{Production-constrained Gravity Model.} In contrast to the globally-constrained gravity model, the production-constrained gravity model necessitates a higher level of granularity in the available information, particularly in terms of the number of trips originating from each region. 
%Thus, the production-constrained gravity model can be used as a solution to enhance the performance of the gravity model, by using the outflow quantity of each region as a constraint. 
As shown in the following equation,
\begin{equation} \label{eq:constrainto}
    \sum_{j=1}^n \hat{T}_{ij} = \sum_{j=1}^n T_{ij} = O_i \quad for \ all \ i ,
\end{equation}
where $O_i$ means the outflow of region $i$. 

The production-constrained gravity model is often preferred due to practical considerations. Researchers may rely on self-reported travel patterns provided by local residents, making this model a more feasible and commonly used approach. 

\textbf{Attraction-constrained Gravity Model.} The attraction-constrained gravity model shares a similar mathematical structure to the production-constrained gravity model, while characterized by the known inflow values. The mathematical expression is shown below,
\begin{equation} \label{eq:constraintd}
    \sum_{i=1}^m \hat{T}_{ij} = \sum_{i=1}^m T_{ij} = D_i \quad for \ all \ i ,
\end{equation}
where $D_i$ denotes the inflow of region $r_i$.

The attraction-constrained mobility model is frequently utilized in estimating non-commuting OD flows~\cite{roy2003spatial,wilson1971family,haynes2020gravity}, such as leisure activities, by leveraging data on visitor volumes at various destinations such as shopping malls, parking lots, and parks. Non-commuting OD flows are an important component of mobility analysis, particularly in urban planning and tourism studies.

\textbf{Doubly-constrained Gravity Model.} The doubly-constrained gravity model represents a further refinement over the two singly-constrained models, as it incorporates constraints on both inflows and outflows. The constraints are combined with Eq. \ref{eq:constrainto} and Eq. \ref{eq:constraintd},
\begin{equation}\label{eq:doubly}
    \begin{array}{ll}
        \begin{aligned}
            \sum_{j=1}^n \hat{T}_{ij} = \sum_{j=1}^n T_{ij} = O_i , \\
            \sum_{i=1}^m \hat{T}_{ij} = \sum_{i=1}^m T_{ij} = D_j .
        \end{aligned}
    \end{array}
\end{equation}

This model plays a pivotal role within the realm of the four-step transportation modeling framework. In real-world scenarios, a variety of facilities can acquire records of inflow and outflow data via monitoring points, including residential neighborhoods, and industrial parks, among others. Consequently, the model demonstrates substantial applicability and relevance in these contexts.

\subsubsection{Theoretical Derivation of Gravity Models Originating From Individual-level} Since the inception of the gravity model, its robustness has been substantiated in a wide array of disciplines. Nevertheless, the majority of these validations predominantly stem from data-driven and empirical perspectives. Numerous academics~\cite{zipf1946p,niedercorn1969economic,fratar1954vehicular} have undertaken rigorous investigations into the fundamental mechanisms of the gravity model from a theoretical perspective, aiming to establish a solid theoretical basis that would enable a more profound comprehension of the gravity model. Given the abundance of literature on the topic, we focus our survey on the most influential and seminal works that have had a significant impact on the field of study. 

\textbf{Zipf's Principle of Least Effort.} The gravity model was first systematically introduced in the seminal work of Zipf in 1946~\cite{zipf1946p}. In this study, Zipf posited that the inter-regional population movement is directly proportional to the product of the population sizes at the origin and destination locations while being inversely proportional to the distance separating them, as illustrated in Eq. \ref{eq:zipf}. Zipf initiated his investigation from the assumption that each individual endeavors to minimize personal effort. By integrating the trade-off between economic paradigms of the localizing economy and big city economy, he was able to derive the gravity model that governs OD flows. The detailed derivation refers to the appendix \ref{apdx:zipf}.

\textbf{Entropy Approach from Statistical Mechanics.} In 1967, Wilson proposed a method based on statistical mechanics, deriving the macroscopic gravity model theory from microscopic mechanisms. The research methodology involves establishing the statistical relationships between microscopic states and macroscopic phenomena. Initially, all microscopic states are considered to have equal probability, and the macroscopic phenomena correspond to the most abundant microscopic states. In this paradigm, microscopic mechanisms help people understand why corresponding macroscopic phenomena occur, and the research on the gravity model follows the same purpose. Correspondingly, in human mobility, the microscopic states represent the choice of origin and destination for each trip, and the macroscopic phenomena manifest as the relationship between population, distance, and OD flows described by the gravity model, as illustrated by Eq. \ref{eq:genG}. The detailed derivation can be obtained in appendix \ref{apdx:entropy}.

\textbf{Minimization of Information Gain from Information Theory.} Roy et al.~\cite{roy2003spatial} provided a comprehensive introduction to the derivation of the gravity model using the minimization of information gain. They updated the model parameters with new data to minimize the information gain from prior probabilities to posterior probabilities. In conjunction with the two types of constraints mentioned above, as shown in Eq. \ref{eq:doubly}, they solved the optimization problem to obtain the expression of the gravity model. The derivation process is shown in Appendix \ref{apdx:InfoGain}.

\textbf{Economic Principles of Utility Maximization.} In 1969, Niedercorn employed the utility theory from economics to derive the gravity law, which led to the traditional gravity principle stating that spatial interaction is directly proportional to regional population and inversely proportional to the distance between the origin and destination~\cite{niedercorn1969economic}. In the realm of economics, the concept of utility refers to a theoretical measure of satisfaction, happiness, or preference that individuals derive from consuming goods or services~\cite{von1947theory,arrow2012social,savage1972foundations}. This abstract notion is employed as a means of quantifying and comparing the relative value or desirability of various options, thus enabling the assessment of decision-making processes. Niedercorn developed a connection between individual utility and personal mobility choices, aiming to optimize the utility gains each person acquires from engaging with other regions through movement. Subsequently, by aggregating these findings to the collective level, he deduced the gravity law as represented by Eq. \ref{eq:zipf}. The complete derivation is given in appendix \ref{apdx:utility}.

In summary, scholars from various fields have assumed different underlying mechanisms and completed the derivation of the gravity model equation based on their own knowledge frameworks and research paradigms. The commonality among these approaches is that they all utilize optimization to derive the gravity model, whether it is through maximizing the macro state with maximum entropy, minimizing costs, or maximizing utility, which may imply the trade-offs between multiple factors.

\subsection{Intervening Opportunities Model} In 1940, Stouffer provided a theory, named Intervening Opportunities Model~(IOM) in human geography~\cite{stouffer1940intervening}, which provides an  explanation for spatial interactions, such as migration and transportation patterns. In parallel with the gravity model~\cite{zipf1946p}, IOM is primarily concerned with elucidating the impact of distance on spatial interactions. 

\subsubsection{Theoretical Framework of Intervening Opportunities Model}
Within the framework of Stouffer's IOM, the relationship between distance and mobility is not characterized by a direct causal link. Rather, this association is mediated by the presence of intervening opportunities. Specifically, in Stouffer's IOM,  individuals making mobility decisions are confronted with two distinct categories of opportunities: direct opportunities and intervening opportunities. Direct opportunities encapsulate the appeal of a specific destination, whereas intervening opportunities delineate the potential disruptions arising from alternative locations that one may encounter on the way to the primary destination. 

% \begin{figure*}
%     \centering
%     \includegraphics[width=0.3\textwidth]{figure/intervening.png}
%     \caption{An illustration of important concepts in the intervening opportunities model.}
%     \label{theory:IOM}
% \end{figure*}

Moreover, Stouffer elucidated the IOM with a rigorous mathematical formulation, capturing the proportional and inverse influences of opportunities and intervening opportunities on the number of individuals migrating to a fixed distance destination~\cite{stouffer1940intervening}. The specific mathematical expression is shown in the following equation,
\begin{equation} \label{eq:IOM}
    \frac{\Delta y}{\Delta s} = \frac{a}{x} \frac{\Delta x}{\Delta s},
\end{equation}
where $\Delta y$ means the number of individuals moving from a certain origin to a circular band of width $\Delta s$ centered at the origin, and at a constant distance $s$ from the origin. To be specific, the distance from the origin to the inner boundary of the circular band is $s - \Delta s /2$ and that of the outer boundary is $s + \Delta s /2$. $x$ denotes the number of intervening opportunities, which is the number of opportunities located within the circular area enclosed by the inner boundary of the circular band. And $\Delta x$ represents the number of opportunities that attract $\Delta y$ travelers to move forward to the circular destination. 
% For a clearer understanding, the above concepts are illustrated in Fig. \ref{theory:IOM}.

\subsubsection{Relationship with Gravity Model}
Derived from Eq. \ref{eq:IOM}, a remarkably distinct inference can be drawn in comparison to the gravity model, suggesting that the role of distance in human mobility might not be as paramount as previously assumed. Furthermore, within the IOM, the aspect of travel cost, which held substantial importance in the gravity model, is now overlooked. In order to integrate the effect of distance into the model, a straightforward method involves formulating intervening opportunities $x$ as a function of distance. This notion can be explicitly illustrated by the subsequent equation,
\begin{equation}
    x = f(s).
\end{equation}
The number of intervening opportunities $x$ is impacted by the area encompassed by the inner boundary of the circular band. More precisely, with the escalation of $s$, $x$ displays a monotonically increasing pattern. As a result, as $s$ enlarges and $x$ increases monotonically, $\Delta y$ demonstrates a monotonically declining tendency, which, to a certain degree, aligns with the inverse association between the volume of population movement and the distance in the gravity model. 

In summary, the principal contribution of the IOM lies in its theoretical framework grounded in sociological methods and paradigms to elucidate the underlying mechanisms of spatial interaction, achieving notable success in this endeavor. Under this theory, OD flows are not simply expected to decrease with increasing distance but are influenced by the distribution of intervening opportunities around the origin. This understanding goes beyond a simple inverse relationship with distance and captures a more essential aspect of the phenomenon. Moreover, investigations into the IOM have persistently thrived, primarily encompassing two dimensions: a) examining and refining the IOM within diverse geographical and social milieus~\cite{haynes2020gravity}, and b) augmenting and enhancing the IOM by amalgamating it with other spatial interaction models~\cite{fotheringham1983some,long1971distance}.

\subsection{Radiation Model} \label{sec:radiation}
In 2012, Simini proposed an alternative physical process analogy to explain the patterns of the population commuting movement in urban areas observed in OD flows~\cite{simini2012universal}. Specifically, Simini likened the job selection decisions of individual actors to radiation emission and absorption processes in physics. In this analogy, the release of particles represented the outflow from each region, and these particles would be absorbed by other regions based on certain criteria, representing the choices of individuals in selecting a job in a particular region. Similar to the research on the gravity model using entropy maximization from statistical mechanics, Simini also adopted the same research paradigm, constructed the individual behavior mechanism, and derived the population mobility patterns at the regional level.

\subsubsection{Theoretical Framework of Radiation Model}

In the modeling of the radiation model, two important concepts are involved, the individuals $\{X\}$, which correspond to the particles in the analogy with the physical radiation process, and the regions. Specifically, we use the index $i$ to represent the origin, i.e., the home location of individuals, and the index $j$ to represent the destination, i.e., the job location. Furthermore, we use $m_i$ and $n_j$ to denote the population size of the origin and destination, respectively. In each region, the number of job opportunities available is determined by $N / N_{job}$, where $N$ represents the population of a region, and $N_{job}$ represents the number of people needed to produce a single job opportunity. It is evident that the number of job opportunities is proportional to the population size. Each job is associated with a corresponding income, and we assume that the distribution of job income follows a certain probability distribution $p(z)$. The income associated with a job opportunity is sampled independently from $p(z)$, based on the number of job opportunities in that region. The absorbance of a region $\mathscr{z}^{(i)}$, which is defined as the maximum job income in that region, represents the attractiveness of that region in terms of attracting individuals from other regions to work there. Each individual also has their own expected income for working in other regions, which is not lower than the highest income they can earn by working in their home region. Therefore, the expected income for each individual is defined as the maximum income among all job opportunities in their home region $\mathscr{z}_X^{(i)}$.

% \begin{figure*}
%     \centering
%     \includegraphics[width=0.3\textwidth]{figure/radiation.png}
%     \caption{An illustration of important concepts in the radiation model.}
%     \label{theory:RM}
% \end{figure*}

Given the above preliminaries, the decision-making process for job selection of each individual analogy with the radiation process can be divided into two steps. First, all particles are released from the home region, and then the released particles are absorbed by other regions, that is, they select to work in a specific region according to a certain rule. The job selection rule is defined as selecting the nearest region with income higher than the expected income $\mathscr{z}_X^{(i)}$. Based on the previous description and rules, we can write the probability formula of the specific working decisions of individuals, that is, the probability of living in the region $i$ and deciding to work in the region $j$. The calculation formula is shown below.
\begin{equation}
    P(1|m_i, n_j, S_{ij}) = \int_0^{\infty} d\mathscr{z} \ P_{m_i}(\mathscr{z}) P_{S_{ij}}(<\mathscr{z}) P_{n_j}(>\mathscr{z}),
\end{equation}
where $\ P_{m_i}(\mathscr{z})$ denotes the probability of highest income of region $i$ through $m_i$ times sampling from $p(z)$, and $S_{ij}$ represents the total population within a circle of radius $r_{ij}$ centered at $i$~($r_{ij}$ is the distance between $i$ and $j$). The computation of $\ P_{m_i}(\mathscr{z})$ is as follow.
\begin{equation}
    \ P_{m_i}(\mathscr{z}) = \frac{d P_{m_i} (<\mathscr{z})}{d \mathscr{z}} = m_i p(<\mathscr{z})^{m_i -1} \frac{dp(<\mathscr{z})}{d \mathscr{z}}.
\end{equation}
As the job incomes are sampled independently, we have that $P_{S_{ij}}(<\mathscr{z}) = p(<\mathscr{z})^{S_{ij}}$ and $P_{n_j}(>\mathscr{z}) = 1 - p(<\mathscr{z})^{n_j}$ obviously. By evaluating the integral of equation x, we derive the following result,
\begin{equation}
    P(1|m_i, n_j, S_ij) = \frac{m_i n_j}{(m_i + S_{ij})(m_i + n_j + S_{ij})},
\end{equation}
which is the probability of individual decision for job selection. By consolidating the aforementioned individual probability distribution to the population level through a binomial distribution, the following expression can be obtained,
\begin{equation}
    P(T_{ij} | m_i, n_j, S_{ij}) = \frac{T_i !}{T_{ij}! (T_i - T_{ij})!} P(1|m_i, n_j, S_ij)^{T_{ij}} (1-P(1|m_i, n_j, S_ij))^{T_{ij}},
\end{equation}
where $T_i$ denotes the outflow of region $i$. The expectation of distribution with the flow information is shown as follows.
\begin{equation} \label{eq:radiation}
    <T_{ij}> = T_i P(1|m_i, n_j, S_{ij}) = T_i \frac{m_i n_j}{(m_i + S_{ij})(m_i + n_j + S_{ij})},
\end{equation}
which is the fundamental equation of the radiation model.

\subsubsection{Relationship With Other Theoretical Models}

The development of the radiation model was proposed as a means to overcome certain constraints inherent in the gravity model~\cite{simini2012universal}. 
Moreover, the radiation model integrates the economic principle of utility maximization, as espoused by the gravity model, by employing a method that extrapolates collective patterns from individual actions within an academic context.
What is more, the relationship between the radiation model and the IOM is more closely and intertwined. Within the context of the radiation model, the allure of the region delineated by $S_{ij}$ can, to some degree, be equated to intervening opportunities~\cite{stouffer1940intervening}. In this model, such opportunities are further specified within distinct circular regions. Moreover, as evidenced by the derived formula, an augmentation in these intervening opportunities results in a reduction in the flow volume of individuals traversing from the origin to the destination.

\subsection{Interconnection Among Three Theoretical Models}

\begin{figure*}
    \centering
    \includegraphics[width=0.85\textwidth]{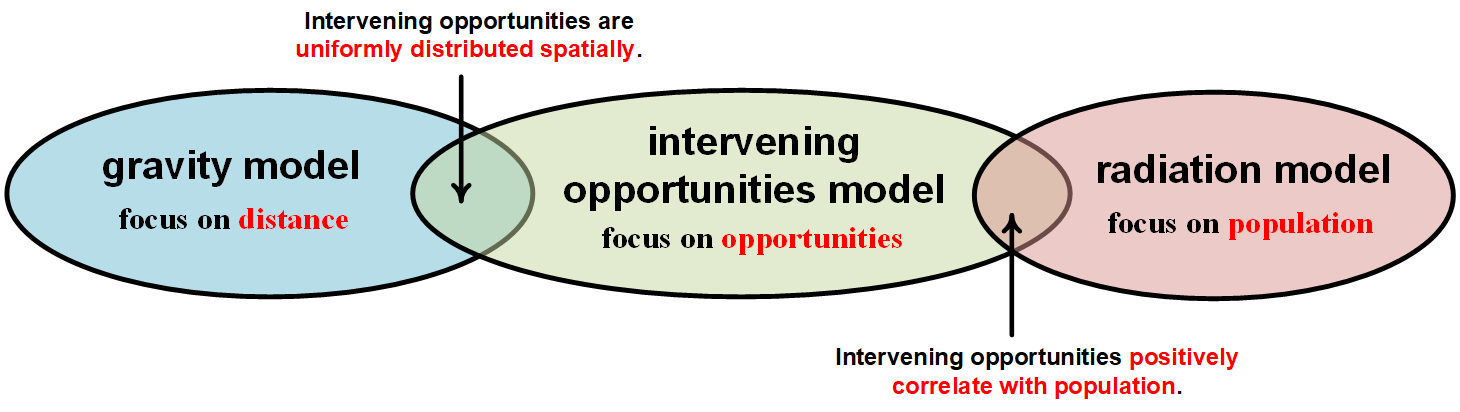}
    \caption{Relationship among the gravity model, intervening opportunities model, and radiation model.}
    \label{3theories}
\end{figure*}

In our summary, the three theoretical models have inherent connections, as illustrated in Fig.~\ref{3theories}. When intervening opportunities are evenly distributed or insignificant, the OD flow exhibits a simple inverse relationship with distance, as expressed in the gravity model. However, when intervening opportunities are proportional to population density, the relationship between OD flow and distance is represented in the radiation model, where population serves as a proxy.

\section{Techniques for Handling Practical Issues Related to OD Flows} \label{sec:tec}

In contrast to the theoretical domains, disciplines such as transportation engineering, urban planning, and computer science concentrate on tackling practical challenges pertaining to OD flows.
In our investigation of research across various domains, we have identified four primary practical problems: OD prediction, OD construction, OD estimation, and OD forecasting. The classification criterion is based on the objectives of the problems in different scenarios and the types of information employed in addressing these issues. Various evaluation approaches are provided in appendix \ref{apdx:metrics}.
% OD prediction encompasses the process of predicting unknown OD flows within a study area by utilizing available information on a subset of known OD flows. This issue garners considerable attention in both transportation and urban planning disciplines. In recent years, the computer science domain has also witnessed endeavors to tackle this challenge through the deployment of advanced machine learning algorithms. The objective of OD construction is to acquire OD flow data for a designated study area where no pre-existing OD flow information is available. OD construction is more frequently encountered in the context of transportation planning and urban planning. OD estimation entails employing observed traffic information, including network flows and vehicular velocities, to deduce the latent OD flows that would yield the corresponding observations upon entering the transportation infrastructure. This approach is also preferred in the transportation field due to the convenience of data and the ability to tailor solutions to specific scenarios. OD forecasting entails the forecasting of future origin-destination flows by integrating real-time observed data with accumulated historical insights. A limited number of transportation service providers, including DiDi and Uber, are able to access taxi origin-destination flow data. These companies place emphasis on anticipating future OD flows in order to optimize vehicle allocation strategies. As a result, there is a considerable body of sophisticated research in the computing domain dedicated to OD flow forecasting.

%%%%
\subsection{Origin-destination Prediction}

% One of the most classical problems that the gravity models and related theories aim to address is inferring unknown OD flows based on the limited information of OD flows between a finite number of locations. 
% In the early stages of research, scholars primarily concentrated on establishing the association between population distribution and OD flows. With the ongoing advancement of the field, increasingly refined data, encompassing demographics, land use patterns, and the distribution of POIs, have been integrated into models to bolster the precision of OD flow predictions. In particular, the models are meticulously constructed to delineate a systematic association encompassing three crucial elements: attributes of the origin, attributes of the destination, and spatial interaction characteristics, such as the distance and travel time, all of which collectively contribute to the corresponding OD flows.

\subsubsection{Preliminaries}

\begin{definition}
    \textbf{Regions.} The urban space is divided into $N$ regions $\mathcal{R} = \{ r_i | i=1,2,...,N \}$. Each region will have a characteristic vector $\mathbf{X}_r$ to represent its urban attributes, which includes socioeconomic and geographic features introduced in Sec \ref{sec:data}.
\end{definition}

\begin{definition}
    \textbf{OD Pairs.} An origin-destination pair denotes a ordered tuple $<r_O, r_D>$, where $r_O$ and $r_D$ represent the origin and destination respectively. We denote the set of all OD pairs of the study area $\mathcal{U}$ as the universal set.
\end{definition}

\textsc{Problem 1.}
\textit{\textbf{OD Prediction.} Given the regional urban characteristics of the city $ \{ \mathbf{X}_r | r \in \mathcal{R} \}  $ and observed OD flows $ \{ f_{ij} | <r_i, r_j> \in \mathcal{X} \} $ between part of OD pairs $\mathcal{X}$ , construct a model to predict the remaining unknown OD flows $ \{ f_{ij} | <r_i, r_j> \notin \mathcal{X} \} $.}

\subsubsection{Methodologies}

There are three categories of techniques and methods for addressing the problem of OD prediction: traditional models, classic machine learning-based models, and neural network-based deep learning models. The summarized description is shown in Table. \ref{Tab:ODpreTech}.

\begin{table}[]
\resizebox{11.5cm}{!}{
\begin{tabular}{|l|c|c|c|l|}
\hline
                                                                                            & Models          & Techniques                                                                  & Required Features                                                                             & Features Type                                                    \\ \hline
\cite{zipf1946p,jin2014location,pourebrahim2018enhancing,karimi2020origin} & Gravity Model   & Physical Model                                                              & Population, distance                                                                          & Numerical                                                        \\ \hline
\cite{stouffer1940intervening,cripps1969empirical,tomazinis1962new}        & IOM             & Social Model                                                                & Opportunities                                                                                 & Numerical                                                        \\ \hline
\cite{simini2012universal,ren2014predicting,lenormand2015influence}        & Radiation Model & Physical Model                                                              & Population                                                                                    & Numerical                                                        \\ \hline
\cite{rodriguez2021origin}                                                 & SVR             & Kernel-based Model                                                          & \begin{tabular}[c]{@{}c@{}}Socioeconomics,\\ distance\end{tabular}                            & Nemerical                                                        \\ \hline
\cite{robinson2018machine}                                                 & GBRT            & Tree-based Model                                                            & Socioeconomics                                                                                & \begin{tabular}[c]{@{}l@{}}Numerical,\\ categorical\end{tabular} \\ \hline
\cite{pourebrahim2019trip}                                                 & Random Forest   & Tree-based Model                                                            & Socioeconomics                                                                                & Numerical                                                        \\ \hline
\cite{jain1996artificial}                                                  & ANN             & Neural Network                                                              & Socioeconomics                                                                                & Numerical                                                        \\ \hline
\cite{yao2020spatial}                                                      & SI-GCN          & Deep Learning                                                               & Socioeconomics                                                                                & \begin{tabular}[c]{@{}l@{}}Numerical,\\ categorical\end{tabular} \\ \hline
\cite{liu2020learning}                                                     & GMEL            & Deep Learning                                                               & Socioeconomics                                                                                & Numerical                                                        \\ \hline
\cite{koca2021origin}                                                      & GCN-MLP         & Deep Learning                                                               & POIs                                                                                          & Numerical                                                        \\ \hline
\cite{cai2022spatial}                                                      & spatialGAT      & Deep Learning                                                               & \begin{tabular}[c]{@{}c@{}}Population,\\ road densities,\\ POIs,\\ railway users\end{tabular} & Numerical                                                        \\ \hline
\cite{yin2023convgcn}                                                      & ConvGCN-RF      & Deep Learning                                                               & \begin{tabular}[c]{@{}c@{}}Population,\\ landuse type\end{tabular}                            & \begin{tabular}[c]{@{}l@{}}Numerical,\\ categorical\end{tabular} \\ \hline
\cite{zeng2022causal}                                                      & SIRI            & \begin{tabular}[c]{@{}c@{}}Deep Learning + \\ Causal Inference\end{tabular} & \begin{tabular}[c]{@{}c@{}}Socioeconomics,\\ POIs\end{tabular}                                & Numerical                                                        \\ \hline
\end{tabular}
}
\caption{A systematic summary on approaches for solving OD prediction problem. The approaches are classified into three types. From top to bottom of the table, there are traditional methods, classic machine learning-based methods and deep learning methods.}
\label{Tab:ODpreTech}
\end{table}

\textbf{Traditional Methods.} The traditional methods including the gravity model, IOM and the radiation model, have been extensively introduced in Sec. \ref{sec:theory}. Recently, the gravity model has been combined with new data to improve the prediction performance. Pourebrahim et al. \cite{pourebrahim2018enhancing} proposed a novel approach to enhance the prediction performance of gravity models by incorporating data from Twitter online check-ins. Jin et al.~\cite{jin2014location} developed an optimized gravity model based on Foursquare check-ins data for their study area. These researches highlighted the potential of LBSN~(Location-based Serviece Network) data. Karimi et al.~\cite{karimi2020origin} proposed a method to improve the accuracy of the gravity model by incorporating traffic counts. Liu et al.~\cite{liu2020urban} proposed a methodology to enhance the gravity model by incorporating land use information. The IOM has undergone development since its inception. One attempt to combine the gravity model and IOM was made by Cripps et al.~\cite{cripps1969empirical}, who employed intervening opportunities as a measure of cost in a gravity model for residential location modeling. Competing opportunities were introduced by Tomazinis et al.~\cite{tomazinis1962new} as an alternative. As for the radiation model, Ren et al.~\cite{ren2014predicting} developed an OD prediction method based on a cost-based generalization of the radiation model and a cost-minimizing algorithm for efficient distribution of the OD flows. Lenormand et al.~\cite{lenormand2015influence} found that the predictive accuracy of the radiation model improved when factors such as population density, income, and education were considered.

\textbf{Classic Machine Learning-based Methods.} This kind of method has been shown in many studies~\cite{robinson2018machine,pourebrahim2019trip,pourebrahim2018enhancing} that outperformed traditional models, demonstrating significantly higher accuracy. Classic machine learning models learn patterns and relationships from a given set of training data to make predictions on unseen data. The classic machine learning models include linear regression, logistic regression, decision trees, Gradient Boosted Regression Trees (GBRT)~\cite{friedman2001greedy}, random forests~\cite{breiman2001random}, support vector machines (SVM), and artificial neural networks (ANN). 
Rodriguez et al.~\cite{rodriguez2021origin} utilized feature engineering to extract significant regional factors that influence OD flows, and conducted a comparison to find that SVR outperformed ANN in terms of accuracy. Sana et al~\cite{sana2018using}. introduced passively collected big data from Google servers to estimate travel demand, comparing ANN, random forest, SVM, and hidden Markov models. 
Robinson et al.~\cite{robinson2018machine} conducted a comparison among ANN, GBRT, and traditional models, and found that machine learning models showed significant advantages over traditional models. 
Pourebrahim et al.~\cite{pourebrahim2019trip} introduced Twitter data and found that the random forest achieved optimal performance. 
Several works~\cite{lenormand2016systematic,pourebrahim2018enhancing,robinson2018machine,pourebrahim2019trip,rodriguez2021origin} have reported the performance of ANN in OD prediction tasks, and in most cases, their performance is inferior to that of SVR and other tree-based machine learning models. 

\textbf{Deep Learning Methods.} Deep learning models have made significant progress recently, particularly in the fields of computer vision (CV)~\cite{he2016deep}, natural language processing (NLP)~\cite{devlin2018bert}, and other areas~\cite{kipf2016semi,li2017deep}. Through a large number of parameters and sophisticated structured design, deep neural networks have shown remarkable fitting and generalization abilities, resulting in high performance in various applications. As deep learning advances, models such as convolutional neural networks~(CNN)~\cite{he2016deep} and graph neural networks~(GNN)~\cite{kipf2016semi}, which are inherently well-suited for extracting urban spatial features, have emerged and have been employed by numerous studies as spatial feature extraction modules~\cite{liu2020learning,yin2023convgcn}. The related works are summarized in Table \ref{Tab:ODpreTech}.
Specifically, Katranji et al.~\cite{katranji2020deep}, Liu et al.~\cite{liu2020learning} and Cai et al.~\cite{cai2022spatial} employ multi-task learning to further improve the prediction accuracy of total OD flows by predicting specific travel pattern OD flows. 
Afandizadeh et al.~\cite{afandizadeh2021hourly} collected extensive ITS data, including automatic number plate recognition~(ANPR) camera data, intersection loop detector data, and smart transit card data, leveraging deep learning techniques to predict travel conditions in metropolitan areas. 
A lot of works~\cite{yao2020spatial,liu2020learning,koca2021origin,cai2022spatial,yin2023convgcn} introduced GNNs to leverage spatial neighborhood similarity to extract features for each region as origin and destination. 
And two works~\cite{cai2022spatial,yin2023convgcn} proposed the hybrid models, which use CNN to process grid-based urban characteristics and combine them with GNNs.
Zeng et al.~\cite{zeng2022causal} proposed a novel approach that combines causal learning, using historical information and urban regional attribute changes to predict future OD flow variations, achieving more robust accuracy on out-of-distribution datasets.

In brief, each of the three kinds of methods has its own advantages and disadvantages. Although traditional methods may not perform well, they offer valuable knowledge and model-building intuition. Traditional machine learning methods achieve decent performance and good explainability in the feature importance. 
%In particular, the ensemble learning strategy such as random forest and GBRT can greatly improve the robustness of the model. 
The deep learning approaches benefit from their flexible design and complex structure to achieve optimal performance. However, due to the large number of parameters, good explainability cannot be realized. In this task, the future direction is to combine the advantages of the three methods and complement each other's strengths. That is, traditional methods can be used to inspire better model designs, such as gravity-inspired GAE~\cite{salha2019gravity}. And specially designed network structures can be used to learn excellent representations and combine them with classic machine learning-based methods to obtain both performance and robustness improvements~\cite{liu2020learning,yin2023convgcn}.

%%%%
\subsection{Origin-destination Construction}

\subsubsection{Preliminaries}

\begin{definition}
    \textbf{OD Matrix.} The OD flows between all regions can be represented in the form of an OD matrix $\mathbf{F}$, where the element $f_{ij}$ at the $i^{th}$ row and $j^{th}$ column indicate the number of people moving from region $r_i$ to region $r_j$. The symbolic representation of the OD matrix can be referred to in Eq. \ref{eq:ODmatrix}.
\end{definition}

\textsc{Problem 2.}
\textit{\textbf{OD Construction.} The OD construction problem aims to construct the complete OD matrix $\mathbf{F}$ for the city based on easily accessible information without any OD flow information available.}

\subsubsection{Methodologies} Relative to OD prediction, OD construction constitutes a considerably more intricate issue. 
% The absence of OD flow data in the target city precludes the development and calibration of models, resulting in potential inaccuracies when applying a model derived from data from other cities. 
Apart from the costly survey-based traditional methods, current solutions to OD construction can be categorized into two types: aggregation of individual trajectories, and building computational models. % The former approach focuses on aggregating and discerning collective OD flow movement patterns from extensive mobile data sources, while the latter primarily employ models to generate OD flow information. 
Following, we designate the former as data-based aggregation methods and the latter as model-based generation methods. A summary of the systematic comparison between these two categories of methods is shown in Table. \ref{Tab:ODconTech}.

\textbf{Data-based Aggregation Methods. } This method emphasizes the aggregation of individual trajectories to ascertain collective OD flow patterns.
The primary intention of this category of methods is to utilize accessible mobility records, such as CDRs and check-ins, thereby replacing the traditional high-cost yearly travel survey approaches. A primary challenge faced in this direction is the need to design different data mining methods according to the distinct characteristics of various data, as shown in Table. \ref{tab:locdata}. 

%According to Table. \ref{tab:locdata}, CDR data is a type of individual location record with coarse spatial granularity and low temporal sampling rates. However, since mobile phones have become indispensable tools in people's daily lives, their user coverage is extensive, and the data collection is vast, which can compensate for the aforementioned shortcomings through massive data over extended time periods. 
There have already been numerous efforts using CDRs as the trajectory information~\cite{caceres2007deriving,calabrese2011estimating,iqbal2014development,alexander2015origin,toole2015path,bachir2019inferring,bonnel2015passive}. 
Duan et al.~\cite{duan2011mobilepulse} explored CDR data for predicting land use types and estimating commuting OD matrices. They identified the home and work locations of each individual based on rules and then aggregated the information to obtain the commuting OD matrix.
Calabrese et al.~\cite{calabrese2011estimating} utilized CDR data from one million users, first obtaining trips based on individual trajectories, and then aggregating these trips to derive the OD flows for the Boston Metropolitan Area.
Iqbal et al.~\cite{iqbal2014development} first identified stay points for all users in the CDR data using spatial clustering, and then aggregated the transitions between stay points for all users to obtain OD flows.
Alexander et al.~\cite{alexander2015origin} further associated trips and activities, and subsequently calculated home-based, work-based, and other types of OD flows based on this relationship.
In addition to the aforementioned studies, there are many similar works or different applications~\cite{wismans2018improving,bonnel2018origin,imai2021origin,bachir2019inferring,heydari2023estimating,luo2020research,tongsinoot2017exploring,gundlegaard2016travel,demissie2016inferring,pourmoradnasseri2019od} focusing on different regions or cities of interest.

There are also some studies that use other types of mobile data to extract OD flows. Some works~\cite{wang2013estimating,pan2006cellular,yang2021detecting,wang2019extracting,fekih2021data} use cellular signaling data to construct OD flows. The difference is that cellular signaling data has a finer spatiotemporal granularity, making it suitable for capturing dynamic OD flows or fine-grained OD flows. Check-ins are also a popular source for extracting OD flow information.
Several existing works~\cite{yang2015origin,osorio2019social,liao2022mobility,janzen2016estimating} have made attempts in this direction. GPS data, as an important source of mobility information, has also been used in numerous studies~\cite{ma2013deriving,wang2019extracting,sadeghinasr2019estimating,munizaga2012estimation,cerqueira2022inference} on OD flows. With the development of ITS, many studies~\cite{gonzalez2020detailed,sari2019high,hamedmoghadam2021automated,cao2021day,hussain2021transit,sun2022flexible,munizaga2012estimation,cerqueira2022inference} utilize data, such as smart card data and station records, etc., to construct OD flow information. 

\begin{table}[]
\resizebox{14cm}{!}{
\begin{tabular}{|c|c|c|c|l|}
\hline
                                                                                            & \begin{tabular}[c]{@{}l@{}}Based on Mobility\\ -related Data\end{tabular} & Data Requirements                                                                         & \multicolumn{1}{c|}{Typical Techniques} & Papers \\ \hline
\multirow{3}{*}{\begin{tabular}[c]{@{}c@{}}Data-based\\ Aggregation\\ Methods\end{tabular}} & \multirow{3}{*}{\checkmark}                                                         & \multirow{3}{*}{\begin{tabular}[c]{@{}c@{}}Individual\\ Mobile Records\end{tabular}}      & Spatial Clustering                       & \cite{calabrese2011estimating,alexander2015origin,toole2015path,bachir2019inferring,bonnel2015passive} \\ \cline{4-5} 
                                                                                            &                                                                           &                                                                                           & Rule-based Filtering                     & \cite{iqbal2014development,alexander2015origin,toole2015path,bachir2019inferring,bonnel2015passive} \\ \cline{4-5} 
                                                                                            &                                                                           &                                                                                           & Statistical Counting                     & \cite{caceres2007deriving,calabrese2011estimating,iqbal2014development,alexander2015origin,toole2015path,bachir2019inferring,bonnel2015passive} \\ \hline
\multirow{3}{*}{\begin{tabular}[c]{@{}c@{}}Model-based\\ Generation\\ Methods\end{tabular}} & \multirow{3}{*}{$\times$}                                                         & \multirow{3}{*}{\begin{tabular}[c]{@{}c@{}}Easily-obtained\\ Auxiliary Data\end{tabular}} & Predictive Models                        & \cite{pourebrahim2018enhancing,liu2020learning,pourebrahim2019trip,zeng2022causal,simini2021deep,rong2021inferring,rong2019deep} \\ \cline{4-5} 
                                                                                            &                                                                           &                                                                                           & Generative Models                       & \cite{rong2023origindestination,rong2023complexityaware} \\ \cline{4-5} 
                                                                                            &                                                                           &                                                                                           & Transfer Learning                        & \cite{rong2023goddag} \\ \hline
\end{tabular}
}
\caption{A systematic comparison between two categories of works on addressing OD construction problem.}
\label{Tab:ODconTech}
\end{table}

\begin{figure*}[t]
    \centering
    \subfigure[Predictive modeling.]{
    \label{fig:genpre}
    \includegraphics[width=0.28\textwidth]{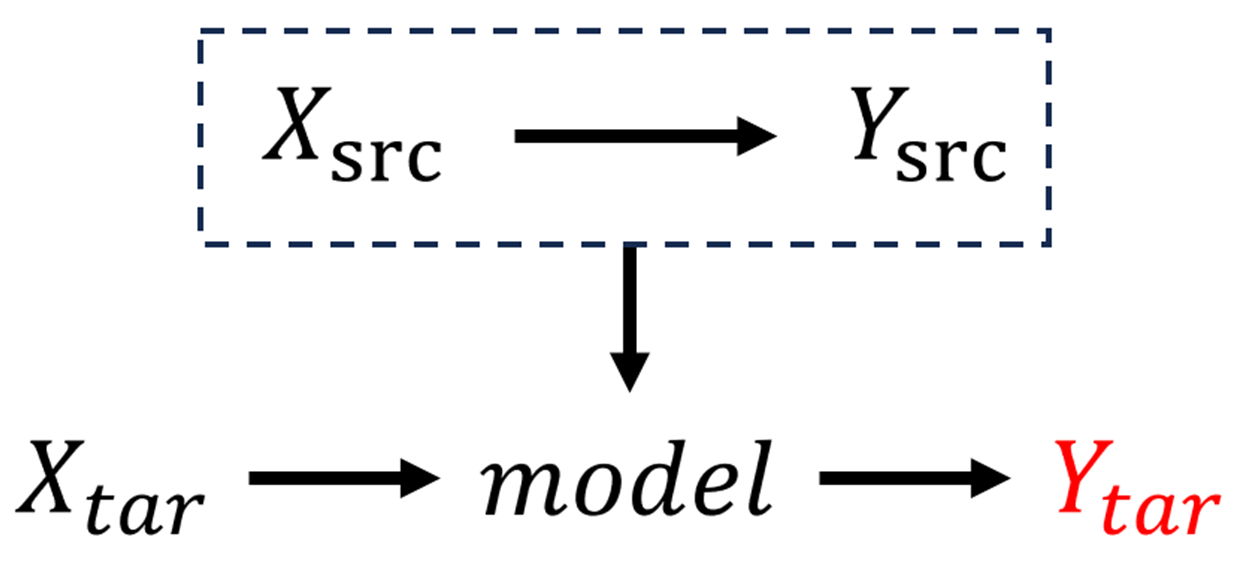}
    }\hfill
    \subfigure[Generative modeling.]{
    \label{fig:gengen}
    \includegraphics[width=0.28\textwidth]{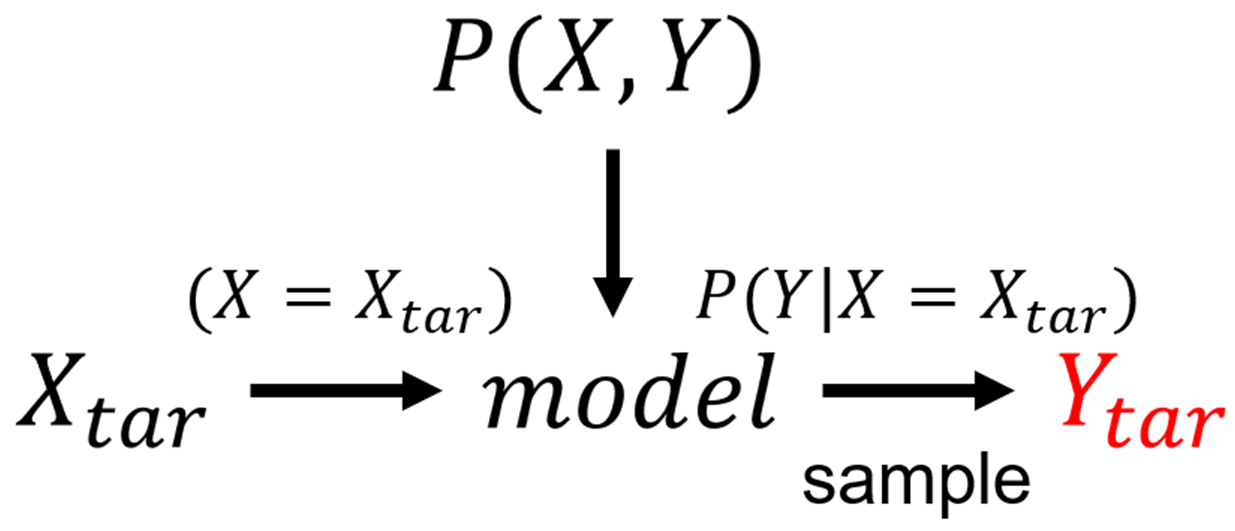}
    }\hfill
    \subfigure[Transfer learning.]{
    \label{fig:gentrans}
    \includegraphics[width=0.36\textwidth]{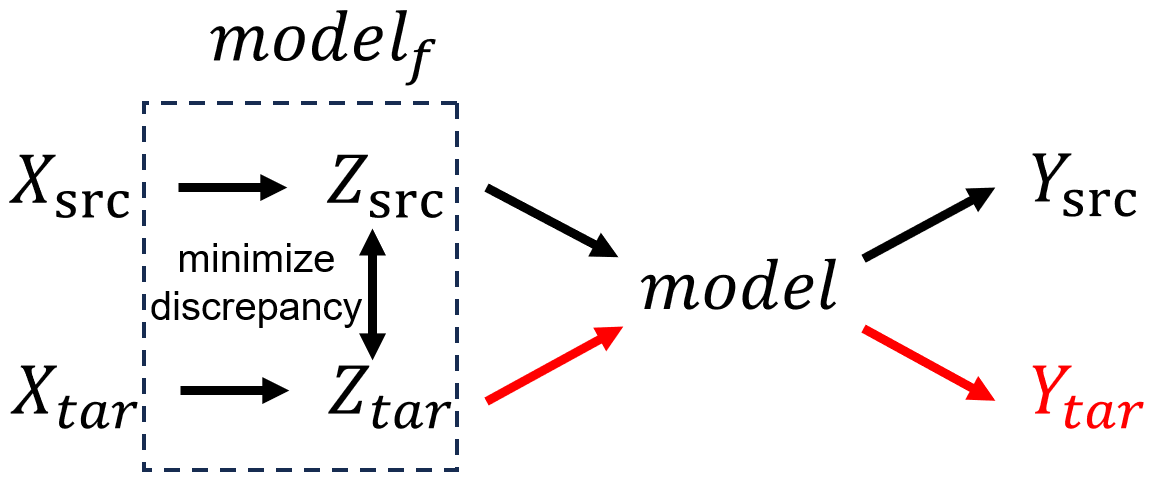}
    }
    \caption{The illustrations of comparing different model-based generation methods for solving OD construction problem.}
    \label{fig:ODgencon}
\end{figure*}

\textbf{Model-based Generation Methods.} Despite the achievements of data-based aggregation methods in addressing the OD construction problem, these approaches entail limitations, including stringent data requirements and potential risks associated with privacy breaches. An alternative approach, akin to OD prediction techniques, entails the development of models that draw upon a city's spatially structured features, thereby generating OD matrices for the target urban area. However, this is a highly futuristic and challenging direction, primarily due to the problem's premise that the target city lacks any OD flow information. Consequently, the model cannot be calibrated through target domain information, resulting in potential errors. These model-based generation works to solve the OD construction problem can be divided into three main categories: predictive models, generative models and transfer learning methods, as shown in Table~\ref{Tab:ODconTech}.

\textbf{Predictive Models.} This kind of method uses predictive models, which target OD flows based on other easily available data. The first is the typical OD prediction techniques~\cite{pourebrahim2018enhancing,liu2020learning,pourebrahim2019trip,zeng2022causal,simini2021deep,rong2021inferring,rong2019deep}, which can be trained directly in a city where OD flow information is obtainable and then predict OD flows between every two regions in the target city, as shown in Fig.~\ref{fig:genpre}. However, there is a limitation that this paradigm will introduce cross-city transfer errors. The OD prediction techniques take the urban characteristics, such as socioeconomics, as the basis of modeling. There are other works that take the dynamic population distribution~\cite{rong2021inferring} and population variations~\cite{rong2019deep} as the input to predict the corresponding OD flows.

\textbf{Generative Models.} Recently, the development of Artificial Intelligence Generated Content~(AIGC) has led to the mainstreaming of the generative models, which has been adopted in OD construction in some works~\cite{rong2023origindestination,rong2023complexityaware}. Rong et al~\cite{rong2023origindestination,rong2023complexityaware} propose to generate OD flow from a network perspective through generative adversarial networks~(GAN)~\cite{goodfellow2020generative} and diffusion models~\cite{ho2020denoising}, which generates the OD flows not only considering the local regional urban attributes but also the complexity~\cite{saberi2017complex,saberi2018complex} of the whole OD matrix of a city, as shown in Fig.~\ref{fig:gengen}.

\textbf{Transfer Learning Methods.} To reduce the error introduced in OD prediction techniques when transferring across cities, another idea is to utilize unsupervised transfer learning methods to improve the transferability of the models. A cutting-edge research direction, known as unsupervised domain adaptation, focuses on training models within a source domain and transferring them to an unlabeled target domain. The underlying idea involves mapping features from both the source and target domains into a common space, aiming to minimize the disparities between them~\cite{rong2023goddag}, as shown in Fig.~\ref{fig:gentrans}. Consequently, this approach allows for the successful application of OD prediction models, initially trained in the source domain, to the target domain. This represents a promising and emerging research direction in the field, with no significant or influential studies available at present.

Until now, data-based aggregation methods have been fully utilized in many scenarios to achieve the task of OD construction. However, the limitation of this kind of method is that it is very demanding in terms of data. Specifically, it should input the location records of a large number of individuals in the city. This information is not easy to be obtained and can disclose the privacy of users. Besides, it has been studied thoroughly because of the simplicity of the technology used. Conversely, model-based generation methods are the future direction, since they require only publicly available auxiliary data. There are already works leveraging transfer learning strategies~\cite{rong2023goddag} and generative models to solve the OD construction problem. With the advent of AIGC, there is a growing trend to use vast amounts of data to learn a universal large model for data generation, which is also the future exploration direction of addressing OD construction tasks.

%%%%
\subsection{Origin-destination Estimation}

\subsubsection{Preliminaries}

\begin{definition}
    \textbf{Temporal OD Flow.} Temporal OD flow refers to the population mobility $\{ f^t_{ij} \}$ between different regions within specified time periods $\{t | t=1,2,...,T\}$. It takes into account both the spatial aspect (i.e., people are moving from the origin $r_i$ and to the destination $r_j$) and the temporal aspect (i.e., when the movement occurs) of the flow.
\end{definition}

\textsc{Problem 3.}
\textit{\textbf{OD Estimation.} Given a set of observed temporal traffic counts or other relevant observations $ \mathcal{T} = \{ x_i^t | t=1,..,T \text{ and } i = 1,...,N \} $ collected at various locations $\{ l_i | i=1,...,N \}$ within a transportation network (such as road segments, intersections, or sensor-equipped locations), the objective is to infer the underlying OD flows $\{f^t_{ij}\}$, i.e., the number of trips between different origin and destination pairs, that generated the observed traffic pattern.}

\subsubsection{Methodologies}
In addressing the OD estimation problem, the adopted strategy entails employing algorithms or models to identify the most like OD flows that correspond to the observed traffic patterns. The fundamental challenge of the origin-destination (OD) matrix estimation problem is that it is severely under-determined. 
Existing approaches to solving the OD estimation problem typically fall into two categories. One consists of traditional methods, which combine traffic models and simulators to find the most likely OD matrix based on optimization. The other involves using data-driven methods, which employ data to construct empirical models and utilize these models to directly or indirectly predict OD flows.

\begin{table}[]
\resizebox{14cm}{!}{
\begin{tabular}{|c|l|c|l|}
\hline
                                       & Objectives & Scenarios          & Papers             \\ \hline
\tabincell{c}{Entropy \\ Maximization} & \tabincell{l}{Maximizing the entropy of the OD Matrix.} & \tabincell{l}{Static OD Estimation} & \cite{mohanty2020dynamic,xie2011maximum,yang1992estimation,van1980most,ait2022value} \\ \hline
\tabincell{c}{Maximum \\ Likelihoold}  & \tabincell{l}{Maximizing the likelihood of dataset} & \tabincell{l}{Static OD Estimation} & \cite{lo1996estimation,jeong2021stochastic,pitombeira2017statistical,dey2020origin,parry2012estimation} \\ \hline
Least Squares                          & \tabincell{l}{Minimizing the errors between models and data.} & \tabincell{l}{Dynamic OD Estimation} & \cite{wang2022transportation,michau2019combining,bell1991estimation,vahidi2022time,cascetta1993dynamic,shen2019spatial} \\ \hline
Primal-Dual                            & \tabincell{l}{Minimizing a function that represents the mismatch \\ between observed and estimated flows}           & \tabincell{l}{Dynamic OD Estimation} & \cite{michau2016primal}            \\ \hline
\tabincell{c}{Bayesian \\ Estimation}  & \tabincell{l}{Incorporate prior knowledge and observed data to \\ estimate the most likely OD flows} & \tabincell{l}{Dynamic OD Estimation} & \cite{pitombeira2020dynamic,pitombeira2016dynamic,croce2021estimation,yu2021bayesian,maher1983inferences}            \\ \hline
\tabincell{c}{Gradient-based \\ Optimization} & \tabincell{l}{Minimizing the mismatch between observations \\ and prediction while the OD flows are the weights} & \tabincell{l}{Dynamic OD Estimation} & \cite{behara2022single,ros2022practical,behara2020novel,frederix2011new,yang2019estimating,cantelmo2015two,cantelmo2017effectiveness,cipriani2011gradient} \\
\tabincell{c}{Genetic \\ Algorithm}    & \tabincell{l}{Minimizing the mismatch between observations and \\ prediction while the OD matrix is the solution space} & \tabincell{l}{Dynamic OD Estimation} & \cite{nigro2018exploiting,ou2019learn,huang2013computational} \\ \hline
\end{tabular}
}
\caption{A summary of traditional methods for addressing origin-destination estimation.}
\label{tab:ODestimTra}
\end{table}

\textbf{Traditional Methods.} This category of methods predominantly frames the OD estimation problem as an optimization task, wherein the objective function to be minimized encompasses five convex components, each representing a constraint or attribute of the transportation problem: adherence to traffic counts, compliance with traffic conservation (Kirchhoff's law), the resemblance of flows originating and culminating in proximate locations, and non-negativity of traffic flows. Consequently, methodologies such as entropy maximization, maximum likelihood estimation, etc. have been adopted to ascertain the most likely OD flows. The techniques employed and their respective features are concisely delineated in Table \ref{tab:ODestimTra}. 
Additionally, there are some works that adopt other methods to solve the optimization problem, such as the Golden Section Search algorithm~\cite{caceres2013inferring}, expectation-maximization~\cite{sun2020origin}, Gauss-Seidel method~\cite{xia2018dimension} and Kalman filter approaches~\cite{lu2015kalman,castiglione2021assignment}.

\textbf{Data-driven Method.} In the field of transportation, researchers have attempted to employ data-driven approaches to construct machine learning models that directly learn the mapping relationship between traffic observation information and OD flows~\cite{song2020dynamic,xiong2023deeplearning}, enabling the estimation of corresponding OD flows based on traffic status. Some works~\cite{tsanakas2023d,krishnakumari2020data,yang2017origin} take an indirect approach, first utilizing neural networks to approximate a computationally complex traffic model and then using that model to accelerate the solution of previous optimization problems. There are even studies~\cite{ma2022estimating,wu2018hierarchical,zheng2021rebuilding} that directly employ neural networks to construct computation graphs, modeling the entire traffic process and subsequently estimating the most likely OD matrix that produces the corresponding traffic observations. %Overall, there is not an abundance of work in this direction while it is a promising and cutting-edge research area.

In summary, traditional methods to estimate the most likely OD flows have efficiency issues and cannot be applied to large-scale city-wide scenarios. Recently, researchers have made progress in improving efficiency by using the data-driven paradigm based on machine learning models instead of traditional traffic models, such as DTA. 
% In the future, combining these two kinds of methods will be a more promising direction. And the main research problem will be to further extend the applicability to a wider range in the urban space.

%%%%
\subsection{Origin-destination Forecasting}

\subsubsection{Preliminaries}

\textsc{Problem 4.}
\textit{\textbf{OD Forecasting.} Given a historical dataset of OD flows $\{ f^t_{ij} | t= 1,2,...,k-1 \}$ over a certain period of time, the objective is to forecast the OD flows for future time periods $\{ 
f^t_{ij} | t=k,k+1,... \}$.}

\subsubsection{Methodologies}

OD forecasting is an autoregressive multivariate time series forecasting problem, similar to classical time series prediction problems, where many classical algorithms can be used as solvers, such as vector auto-regression~(VAR)~\cite{hamilton2020time}, support vector regression~(SVR)~\cite{smola2004tutorial}, and auto-regressive integrated moving average~(ARIMA)~\cite{box2015time}. With the development of deep learning, many computer science researchers have introduced advanced spatiotemporal sequence prediction models~\cite{yan2018spatial,guo2019attention,geng2019spatiotemporal} into the field of OD forecasting, significantly improving prediction performance. And there are advanced deep learning methods as shown in Table~\ref{tab:ODfore}.
%Graph structures are naturally introduced to construct associations between multiple spatial location time series, and various advanced graph neural networks~\cite{kipf2016semi,velivckovic2017graph,xu2018powerful} are used to model the dependencies between different spatial location sequences. Time series feature modeling has many advanced deep learning models to choose from, ranging from classic RNN~\cite{schuster1997bidirectional}, LSTM~\cite{hochreiter1997long}, GRU~\cite{chung2014empirical}, TCN~\cite{lea2017temporal}, to the current advanced transformer with self-attention mechanisms, all of which have been chosen by existing works as time series feature processing modules~\cite{vaswani2017attention}. 
%In summary, the general approach of existing OD forecasting works is to further improve various modules within a modular framework, introducing a variety of novel designs to enhance the performance of OD forecasting in specific scenarios.

\begin{table}[]
\resizebox{14cm}{!}{
\begin{tabular}{|c|l|c|c|c|}
\hline
Models     & Spatial Topology Construction                                                                                                                                                                                  & Spatial Feature Extraction            & Temporal Modeling                   & Learning Strategies                                                                   \\ \hline
gMHC-STA~\cite{bhanu2022graph}   & \begin{tabular}[c]{@{}l@{}}region-pairs as nodes\\ full-connected graph\end{tabular}                                                                                                                           & GCN + spatial attention               & self-attention                      & MSELoss                                                                                \\ \hline
ST-VGCN~\cite{yang2022spatiotemporal}    & \begin{tabular}[c]{@{}l@{}}region-pairs as nodes\\ Pearson correlation graph\end{tabular}                                                                                                                      & GCN + gated mechanism                 & GRU                                 & MSELoss                                                                                \\ \hline
MVPF~\cite{zheng2022metro}       & \begin{tabular}[c]{@{}l@{}}stations as nodes\\ distance-based graph\end{tabular}                                                                                                                               & GAT                                   & GRU                                 & MSELoss                                                                      \\ \hline
Hex D-GCN~\cite{yang2021origin}  & \begin{tabular}[c]{@{}l@{}}hexagonal grids as nodes\\ taxi path-based dynamic graph\end{tabular}                                                                                                            & GCN                                   & GRU                                 & MSELoss                                                                                \\ \hline
CWGAN-GP~\cite{li2022network}   & OD matrix as an image                                                                                                                                                                                          & CNN                                   & CNN                                 & \begin{tabular}[c]{@{}l@{}}GAN-based Training\\ condition on histories\end{tabular} \\ \hline
SEHNN~\cite{zhao2022station}      & \begin{tabular}[c]{@{}l@{}}stations as nodes\\ geo-adjacency graph\end{tabular}                                                                                                                                & GCN                                   & LSTM                                & VAE-based Training                                                                  \\ \hline
HC-LSTM~\cite{miao2022deep}    & \begin{tabular}[c]{@{}l@{}}grids as nodes\\ OD flow-based graph\\ in/out flow as an image\\ OD matrix as an image\end{tabular}                                                                                  & CNN + GCN                             & LSTM                                & MSELoss                                                                                \\ \hline
Gallat~\cite{wang2021gallat}     & \begin{tabular}[c]{@{}l@{}}regions as nodes\\ OD flow-based graph\\ distance-based graph\end{tabular}                                                                                                          & spaital attention                     & temporal attention                  & MSELoss                                                                                \\ \hline
ST-GDL~\cite{zou2021long}     & \begin{tabular}[c]{@{}l@{}}regions as nodes\\ distance-based graph\end{tabular}                                                                                                                                & CNN + GCN                             & CNN                                 & MSELoss                                                                  \\ \hline
PGCM~\cite{li2020graph}       & \begin{tabular}[c]{@{}l@{}}region pairs as nodes\\ OD flow-based graph\end{tabular}                                                                                                                      & GCN + gated mechanism                             & none                                & probabilistic inference with Monte Carlo                                            \\ \hline
MF-ResNet~\cite{he2022short}  & OD matrix as an image                                                                                                                                                                                          & CNN                                   & none                                & MSELoss                                                                   \\ \hline
TS-STN~\cite{jiang2022deep}     & \begin{tabular}[c]{@{}l@{}}stations as nodes\\ OD flow-based graph\end{tabular}                                                                                                                            & \tabincell{c}{temporally shifted \\ graph convolution}    & LSTM + attention                    & Partially MSELoss                                                           \\ \hline
ODP-URS~\cite{noursalehi2021dynamic}          & OD matrix as an image                                                                                                                                                                                          & CNN                                   & LSTM                                & MSELoss                                                                  \\ \hline
DMGC-GAN~\cite{huang2022gan}   & \begin{tabular}[c]{@{}l@{}}regions as nodes\\ geo-adjacency graph\\ OD flow-based graph\\ in/out flow-based graph\end{tabular}                                                                           & GCN                                   & GCN + GRU                           & GAN-based training                                                                  \\ \hline
DNEAT~\cite{zhang2021dneat}      & \begin{tabular}[c]{@{}l@{}}regions as nodes\\ geo-adjacency graph\\ OD flow-based graph\end{tabular}                                                                                                           & attention                             & attention                           & MSELoss                                             \\ \hline
CAS-CNN~\cite{zhang2021short}    & OD matrix as an image                                                                                                                                                                                          & CNN                                   & channel-wise attention              & masked loss function                                                                \\ \hline
ST-ED-RMGC~\cite{ke2021predicting} & \begin{tabular}[c]{@{}l@{}}region pairs as nodes\\ fully-connected graph\\ geo-adjacency graph\\ POI-based graph\\ distance-based graph\\ OD flow-based graph\end{tabular} & GCN                                   & LSTM                                & MSELoss                                                                                \\ \hline
HSTN~\cite{chen2022origin}       & \begin{tabular}[c]{@{}l@{}}regions as nodes\\ geo-adjacency graph\\ node pattern similarity graph\end{tabular}                                                                                                 & GCN                                   & GRU+Seq2Seq                         & MSELoss                                                                                \\ \hline
BGARN~\cite{shen2022baselined}      & \begin{tabular}[c]{@{}l@{}}grid clusters as nodes\\ distance-based graph\\ OD flow-based graph\end{tabular}                                                                                                   & GCN+attention                         & LSTM                                & MSELoss                                                                                \\ \hline
HMOD~\cite{zhang2022dynamic}       & \begin{tabular}[c]{@{}l@{}}regions as nodes\\ OD flow-based graph\end{tabular}                                                                                                                                 & \tabincell{c}{random walk \\ for embedding}             & GRU                                 & MSELoss                                                                                \\ \hline
STHAN~\cite{ling2023sthan}      & \begin{tabular}[c]{@{}l@{}}regions as nodes\\ geo-adjacency graph\\ POI-based graph\\ OD flow-based graph\end{tabular}                                                                                & \tabincell{c}{convolution by \\ meta-paths + attention}  & GRU                                 & MSELoss                                                                 \\ \hline
ODformer~\cite{huang2023odformer}   & regions as nodes                                                                                                                                                                                               & \tabincell{c}{2D-GCN within \\ Transformer}             & none                                & MSELoss                                                                                \\ \hline
CMOD~\cite{han2022continuous}       & \begin{tabular}[c]{@{}l@{}}stations as nodes\\ passengers as edges\end{tabular}                                                                                                                                & \tabincell{c}{multi-level inform \\ -ation aggregation}   & \tabincell{c}{multi-level inform \\ -ation aggregation} & continous time forecasting                                                          \\ \hline
HIAM~\cite{liu2022online}       & \begin{tabular}[c]{@{}l@{}}stations as nodes\\ railway-based graph\end{tabular}                                                                                                                          & GCGRU                                 & GCGRU + Transformer                 & online forecasting                                                                  \\ \hline
DAGNN~\cite{dapeng2021dynamic}      & \begin{tabular}[c]{@{}l@{}}regions as nodes\\ fully-connected graph\end{tabular}                                                                                                                               & subgraph + graph convolution          & TCN                                 & MSELoss                                                                                \\ \hline
GEML~\cite{wang2019origin}       & \begin{tabular}[c]{@{}l@{}}regions as nodes\\ geo-adjacency graph\\ POI-based graph\end{tabular}                                                                                                          & GCN                                   & LSTM                                & multi-task learning                                                                 \\ \hline
MPGCN~\cite{shi2020predicting}      & \begin{tabular}[c]{@{}l@{}}regions as nodes\\ distance-based graph\\ POI-based graph\\ OD flow-based graph\end{tabular}                                                                         & 2DGCN                                 & LSTM                                & MSELoss                                                                                \\ \hline
GDCF~\cite{han2023generic}       & \begin{tabular}[c]{@{}l@{}}regions or stations as nodes \\ clusters as nodes \\ distance-based graph\\ OD flow-based graph\end{tabular}                                                                         & weighted aggregation                                 & \tabincell{c}{continuous \\ updating \\ mechanism}            & Normalized MSELoss                                                                                \\ \hline
\end{tabular}
}
\caption{A summary comparison of the works on the origin-destination forecasting problem.}
\label{tab:ODfore}
\end{table}

A comprehensive summary of existing OD forecasting-related works is presented in Table \ref{tab:ODfore}. From this, we can see that GNN is a commonly used approach for extracting spatial features. Specifically, some works use GCN~\cite{kipf2016semi}, while others employ GAT~\cite{velivckovic2017graph} or 2DGCN~\cite{shi2020predicting} to address spatial modeling. There are roughly three ways to construct graphs, including using spatial adjacency relationships~\cite{zheng2022metro,wang2021gallat,zou2021long,huang2022gan,ke2021predicting,chen2022origin,shen2022baselined}, urban functional similarity~\cite{huang2022gan,ke2021predicting,ling2023sthan,liu2020learning,shi2020predicting}, historical time series pattern similarity~\cite{yang2022spatiotemporal,li2020graph,jiang2022deep,huang2022gan,ke2021predicting,chen2022origin,ling2023sthan,wang2019origin,shi2020predicting}, and OD flows correlation~\cite{miao2022deep,wang2021gallat,huang2022gan,zhang2021dneat,shen2022baselined,zhang2022dynamic}. For handling temporal information, common sequence modeling models such as LSTM~\cite{zhao2022station,miao2022deep,jiang2022deep,noursalehi2021dynamic,ke2021predicting,shen2022baselined,wang2019origin,shi2020predicting}, GRU~\cite{yang2022spatiotemporal,zheng2022metro,yang2021origin,huang2022gan,chen2022origin,zhang2022dynamic,ling2023sthan,liu2022online}, TCN~\cite{dapeng2021dynamic}, and self-attention~\cite{bhanu2022graph,huang2023odformer,liu2022online} are employed. Most works train models using the error between the ground truth and the predictions in the training set, while a few works utilize GAN-based loss~\cite{li2022network,huang2022gan} or VAE-based loss~\cite{zhao2022station} for model training. It is worth noting that sparsity is a unique attribute of OD flow that most works consider separately, and they use gated mechanisms~\cite{yang2022spatiotemporal,li2020graph} to model sparsity, which improves forecasting accuracy.

In the future, there are two main avenues to improve the accuracy of OD forecasting. Firstly, in terms of data, the focus extends beyond historical OD flow information to include POIs, traffic condition data, and other relevant factors. Incorporating such information can enhance the predictive accuracy of OD forecasting. Secondly, in terms of modeling, advanced graph-based modeling techniques and spatiotemporal neural networks, such as graph transformers~\cite{yun2019graph}, can be further employed. These advanced modeling approaches enable the exploration of complex spatial and temporal relationships within OD flows.

\subsection{Comparative Overview of the Four Problems and Their Benchmarks}

\begin{table}[]
\resizebox{14cm}{!}{
\begin{tabular}{|l|c|l|c|l|}
\hline
                                 & Scenarios    & \multicolumn{1}{c|}{\begin{tabular}[c]{@{}c@{}}Available \\ Mobility \\ Information\end{tabular}} & Input                 & \multicolumn{1}{c|}{Why}                                                                                                              \\ \hline
OD Prediction                    & Intra-city   & Partial OD Flows                                                                                  & Urban Characteristics & \begin{tabular}[c]{@{}l@{}}The spatial heterogeneity in the distribution\\ of urban functions drives human mobility.\end{tabular}     \\ \hline
\multirow{2}{*}{OD Construction} & Cross-cities & \begin{tabular}[c]{@{}l@{}}OD Flows Outside \\ The Target City\end{tabular}                       & Urban Characteristics & \begin{tabular}[c]{@{}l@{}}The spatial heterogeneity in the distribution\\ of urban functions drives human mobility.\end{tabular}     \\ \cline{2-5} 
                                 & Intra-city   & \begin{tabular}[c]{@{}l@{}}Mobility Trajectories \\ of Individuals\end{tabular}                   & Trajectories          & \begin{tabular}[c]{@{}l@{}}OD flows are statistical representations of \\ individual trajectories at the regional level.\end{tabular} \\ \hline
OD Estimation                    & Intra-city   & None                                                                                              & Traffic conditions    & Human mobility impacts traffic conditions.                                                                                            \\ \hline
OD Forecasting                   & Intra-city   & \begin{tabular}[c]{@{}l@{}}Historical OD Flows \\ Between All Regions\end{tabular}                & Historical OD Flows   & \begin{tabular}[c]{@{}l@{}}Perform autoregressive forecasting based \\ on its own patterns.\end{tabular}                              \\ \hline
\end{tabular}
}
\caption{A comparison among basic characteristics of the four problems.}
\label{tab:fourpsum}
\end{table}

We will now provide a comparative overview to summarize the characteristics of the four problems. As shown in Table~\ref{tab:fourpsum}, different scenarios have varying degrees of access to OD flow-related information, and algorithms can be designed accordingly based on the available level of OD flow information. Furthermore, with different OD flow information known, we can generate complete OD flow information by considering different facets.

\begin{figure*}
    \centering
    \includegraphics[width=0.8\textwidth]{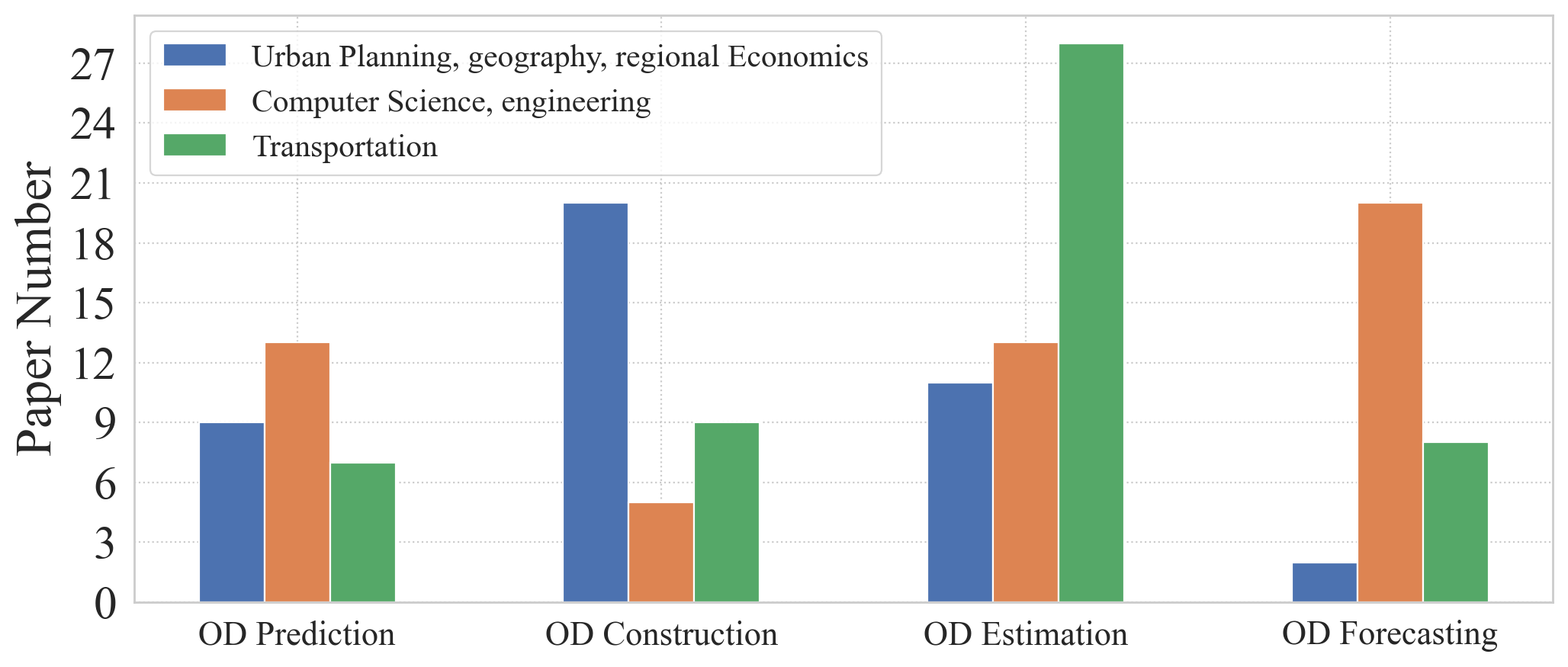}
    \caption{A comparison of paper numbers of different fields working on solving the four problems.}
    \label{fig:fourpradar}
\end{figure*}

% \begin{figure*}
% \centering
% \subfigure[OD prediction.]{ % 
%     \label{four:pre}
%     \includegraphics[width=0.35\textwidth]{figure/pre.png}
% }\hfill
% \subfigure[OD construction.]{ % 
%     \label{four:con}
%     \includegraphics[width=0.35\textwidth]{figure/con.png}
% }\hfill
% \subfigure[OD estimation.]{ % 
%     \label{four:est}
%     \includegraphics[width=0.35\textwidth]{figure/est.png}
% }\hfill
% \subfigure[OD forecasting.]{ % 
%     \label{four:for}
%     \includegraphics[width=0.35 \textwidth]{figure/for.png}
% }
% \caption{A comparison of paper numbers of different fields working on solving the four problems.}
% \label{fig:fourpradar}
% \end{figure*}

We present the contributions of different disciplines to the four problems in Fig.~\ref{fig:fourpradar}, using a bar chart for visualization. Three types of disciplines are included. From the chart, we can observe that OD prediction and OD construction receive more attention from fundamental theoretical disciplines and macro-scale level disciplines. OD estimation is primarily studied in the transportation field, while OD forecasting is mainly researched in computer science and system engineering. 

Especially, to facilitate future research and promote the advancement of related studies, we have collected~\cite{antoniou2016towards} and developed  benchmarks for the aforementioned four problems, and they have been made publicly available at https://github.com/tsinghua-fib-lab/OD\_benckmark.

\section{Applications} \label{sec:app}
\subsection{Urban Planning}

In urban planning, understanding human mobility patterns has become a critical aspect for the effective design, management, and evaluation of urban spaces~\cite{song2010modelling,zheng2014urban}. OD flows and population mobility models play a pivotal role in capturing the complex interactions between people, transportation systems, and the built environment. These models provide valuable insights into the daily movements of residents within and between urban areas, revealing underlying spatial and temporal patterns that help planners make informed decisions on various aspects of urban development. Some works are introduced below.

%Urban planning is a field that focuses on the long-term development of cities, with the OD flow information used in this context typically being oriented towards long-term flows, such as daily OD flows or commuting OD flows. These flows are closely related to urban structure and are often utilized to evaluate planning policies or identify flaws. 
\begin{packed_itemize}
    % \item Anas et al.~\cite{anas1998urban} provides an overview of the literature on urban spatial structure and its relationship with transportation. It discusses the application of existing OD flows and population mobility models in understanding the spatial structure of cities, as well as the implications for urban planning and policy-making.
    % \item Cervero et al.~\cite{cervero2010effects} examine the effects of built environments on vehicle miles traveled (VMT) using population mobility from 370 US urbanized areas. 
    % The study provides insights into the relationship between urban form and travel behavior, which can guide urban planning decisions.
    \item Zhong et al.~\cite{zhong2014detecting} use existing OD flows to detect the dynamics of urban structure through spatial network analysis. 
    % They identify key urban centers and their evolution over time, providing insights for urban planning and development.
    % \item Louail et al.~\cite{louail2014mobile} uses existing mobile phone data to derive OD flows and analyze the spatial structure of cities. The authors identify universal patterns in urban mobility that can guide urban planning and design.
    % \item Kim et al.~\cite{kim2005life}  explores the role of life cycle and environmental factors in selecting residential and job locations using existing OD matrix data. The authors investigate the influence of various factors on travel demand and provide insights for urban planning and policy development.
    \item Yuan et al.~\cite{yuan2012discovering} investigate regions of different functions in a city using human mobility and points of interest (POIs) data. They leverage existing OD matrix data and population mobility models to identify distinct urban functional regions, which can guide urban planning and policy-making.
    % \item Barthelemy et al.~\cite{barthelemy2009co} develop a simple model of city formation that co-evolves density and topology. They utilize existing mobility models to analyze the structure of urban road networks and provide insights for urban planning and road network design.
    % \item Wegener et al.~\cite{wegener2004land} reviews the state of the art in land-use transport interaction modeling, with an emphasis on integrating OD matrix data and population mobility models. The authors discuss various models and their applications in urban planning and policy analysis.
    % \item Anas et al.~\cite{anas1996general} develop a general equilibrium model of polycentric urban land use with endogenous congestion and job agglomeration. They use existing OD flows related mobility models to analyze the spatial distribution of jobs and residences and provide insights for urban planning and congestion management.
    \item Sun et al.~\cite{sun2013understanding} use existing population mobility models to examine daily encounters in metropolitan areas. They analyze various urban factors that influence the frequency and distribution of encounters, providing insights for urban planning and policy making.
\end{packed_itemize}

\subsection{Traffic Management and Analysis}

OD flows play a crucial role in transportation management and analysis as a fundamental input for numerous transportation studies. By capturing the movement of people between various regions within a given city, OD flows provide essential insights into the demand for transportation services and travel patterns~\cite{su2024metrognn}. 
%In the field of transportation management, the analysis of OD flow data enables the optimization of transportation network design and real-time operation, such as the scheduling of public transit services, resource allocation, and the implementation of dynamic pricing strategies. 
By examining OD flows, transportation authorities can identify high-demand corridors, anticipate congestion, and develop targeted interventions to improve overall system performance and user experience. Furthermore, OD flows serve as a valuable tool for transportation analysis by facilitating the evaluation of existing infrastructure and transportation policies, as well as the forecasting of future demand. This empowers planners and policymakers to assess the effectiveness of various measures, such as the introduction of new transit lines, congestion pricing, or changes in transport policies, and to identify potential bottlenecks or areas in need of investment. 

There are examples of applying OD flow information for transportation applications.

\begin{packed_itemize}
    \item Cats et al.~\cite{cats2014dynamic} apply OD flows and population mobility models to analyze the dynamic vulnerability of public transport networks. They evaluate the impact of service disruptions and examine the substitution effects and mitigation measures that can be implemented to minimize the adverse effects on passengers.
    \item Lattman et al.~\cite{lattman2016development} introduces the Perceived Accessibility Scale (PAC) to evaluate passengers' perceptions of public transport accessibility. The researchers utilize OD matrix data to assess the spatial distribution of travel demand and analyze how these factors influence the perceived accessibility of public transportation services.
    %\item Cervero et al.~\cite{cervero1997travel} discusses the relationship between urban form and travel demand, using OD flows with mobility models to analyze the impact of density, diversity, and design on transportation patterns. The authors provide valuable insights for urban planners and policymakers in creating more sustainable and efficient urban environments.
    \item Pereira et al.~\cite{pereira2015using} explore the use of web data to predict public transport arrivals during special events. They utilize existing OD matrix data to create accurate predictions, which can help urban planners optimize public transportation services.
\end{packed_itemize}

\subsection{Epidemic Control}

In the context of epidemic control, OD flow data is utilized to inform public health interventions, such as the implementation of travel restrictions, quarantine measures, and targeted vaccination campaigns~\cite{balcan2009multiscale,kraemer2020effect}. By analyzing OD flows, public health officials can identify high-risk areas, anticipate potential outbreaks, and develop targeted strategies to mitigate the spread of infectious diseases and protect vulnerable populations~\cite{kraemer2020effect}. Furthermore, OD flows can be integrated with epidemiological models to better predict the spatial and temporal dynamics of disease transmission. This enables health authorities to assess the effectiveness of various control measures, such as social distancing policies, contact tracing efforts, and changes in healthcare infrastructure, and to allocate resources efficiently to the areas most in need~\cite{kraemer2020effect}. Some works are as follows.

% Investigating epidemic transmission in relation to population mobility has consistently been a prominent area of interest within the academic community.

\begin{packed_itemize}
    \item Balcan et al.~\cite{balcan2009multiscale} investigates the role of multiscale mobility networks in the spatial spreading of infectious diseases. The authors use an OD matrix derived from census data and airline transportation data to model the spreading patterns of diseases, providing valuable insights for epidemic control and public health interventions.
    % \item Wesolowski et al.~\cite{wesolowski2014commentary} discuss the potential of using mobile network data to inform epidemic control strategies during the Ebola outbreak. They propose using population mobility models derived from mobile phone data to predict disease spread and inform targeted interventions.
    % \item Yang et al.~\cite{yang2015transmission} analyzes the transmission network of the 2014-2015 Ebola epidemic in Sierra Leone. The authors use an OD matrix based on mobile phone call data to model disease transmission patterns and identify high-risk areas, which can inform epidemic control measures and resource allocation.
    % \item Stoddard et al.~\cite{stoddard2009role} examines the role of human movement in the transmission of vector-borne pathogens. The authors use population mobility data derived from GPS tracking devices and an OD matrix to model disease transmission dynamics, providing insights for targeted interventions and epidemic control.
    \item Zhang et al.~\cite{zhang2017spread} model the spread of the Zika virus in the Americas using an OD matrix based on airline transportation data and population mobility models. The study provides valuable information for public health interventions and epidemic control strategies.
    % \item Wesolowski et al.~\cite{wesolowski2012quantifying} use mobile phone data to quantify the impact of human mobility on malaria transmission. By incorporating existing OD matrix and population mobility models, they are able to identify areas of high transmission risk, informing targeted intervention strategies for epidemic control.
    \item Jia et al.~\cite{jia2020population} investigates the role of population flow in driving the spatio-temporal distribution of COVID-19 in China. The authors use existing OD flows and population mobility data to model the spread of the virus, providing valuable insights for epidemic control and public health policy.
    % \item Finger et al.~\cite{finger2016mobile} leverage mobile phone data to examine the role of mass gatherings in the spreading of cholera outbreaks. They incorporate existing OD flow information to predict the spread of the disease, informing targeted intervention strategies for epidemic control.
    \item Huang et al.~\cite{huang2020rapid} presents a contact-tracing investigation of a cluster of COVID-19 cases among young people aged 16-23 years. The authors use existing OD flows to analyze the transmission dynamics of the virus, providing important information for epidemic control and public health policy.
    % \item Kraemer et al.~\cite{kraemer2020effect} investigates the effect of human mobility and control measures on the COVID-19 epidemic in China. The authors use mobility data to assess the impact of travel restrictions and other interventions on the epidemic's spread, providing valuable insights for epidemic control.
\end{packed_itemize}

\section{Open Challenges and Future Directions}\label{sec:future}
\subsection{Challenges}

 OD flows play a pivotal role in capturing human mobility patterns and bears substantial applications and functions for various disciplines, such as urban planning, transportation management, and epidemic control. Nonetheless, several outstanding challenges in this domain warrant attention to fully harness its potential.

\begin{packed_itemize}
    \item \textbf{Explainable Modeling based on Various Urban Features.} Traditional theory-based OD flow modeling has strong interpretability~\cite{stouffer1940intervening,simini2012universal,haynes2020gravity} but, due to its limited consideration of factors and model simplicity, it lacks accuracy. Advanced data-driven approaches can employ complex models to consider numerous intricate factors, offering higher accuracy but lower explainability~\cite{liu2020learning,yin2023convgcn}. For OD flow modeling to provide deeper insights into population mobility and better assist decision-makers, both model effectiveness and explainability must be considered. However, this is a highly challenging task.
    % \item \textbf{Modeling the Complexity of OD Matrix.} Existing research has highlighted the need to consider complexity in urban modeling~\cite{caldarelli2023role}.  In particular, it is necessary to use a micro-level underlying mechanism to more profoundly understand the emergence of macro-level phenomena in population mobility, as manifested in OD flows. These flows, when viewed from a network perspective, exhibit scaling behaviors~\cite{saberi2017complex,saberi2018complex}. However, discovering and modeling such underlying mechanisms is highly challenging.
    \item \textbf{Data Quality and Availability.}  a) Obtaining accurate, high-resolution, and up-to-date OD flow data can be difficult, particularly in regions with limited data infrastructure or privacy concerns. Additionally, data from different sources may have varying levels of quality and granularity, making it challenging to integrate and analyze them. b) Gaining access to data from specialized scenarios, such as disaster contexts, poses considerable difficulties. 
    \item \textbf{Scalability of Models and Algorithms for High-dimensional OD Matrix Data.} OD matrices can exhibit a regional square magnitude, encompassing millions of dimensions, as they represent the origin-destination relationships between all possible pairs of regions within a given area. This high-dimensional nature of OD matrices presents significant computational and analytical challenges. 
\end{packed_itemize}

In order to address this challenge, researchers need to explore and develop innovative modeling approaches, such as machine learning, deep learning, and complex network analysis, which can effectively capture and leverage the high-ordered spatiotemporal dependencies inherent in dynamic OD matrices. 

\subsection{Future Directions}

In light of the prevailing challenges outlined above, we propose two promising avenues for addressing the obtaining problem of OD flow information from perspectives of model design and introducing new algorithms.

\subsubsection{Theory-driven Advanced Model Design}

Based on the aforementioned theoretical research on OD flows, we can summarize three promising future directions that can be guided by theory.

\begin{packed_itemize}
    \item \textbf{Data and Knowledge Jointly-driven Modeling.} Pure data-driven methods emphasize accuracy while neglecting the explainability of the model, whereas purely knowledge-based modeling methods have strong explainability. The incorporation of physical knowledge into neural network design~\cite{karniadakis2021physics,zhang2022physics}, aims to enhance the modeling, thereby boosting both generalizability and explainability. The use of machine learning methods to rediscover and discover new knowledge, effectively employing these techniques not just for predictions, but also as tools for knowledge extraction~\cite{cranmer2020discovering,shi2023learning,sun2022symbolic,cornelio2023combining}.
    % \item \textbf{Data-driven Complexity Modeling.} The nature of population mobility in cities is the result of multiple interactions among urban spaces, individuals, and between individuals~\cite{caldarelli2023role}. These interactions can often be represented in the form of networks, and the rapid development of graph modeling techniques~\cite{velivckovic2017graph,vignac2022digress} in deep learning currently offers powerful tools for studying the complex science of population mobility underlying OD flows. Starting from the data, these methods capture the patterns of networks through abstract representations.
    \item \textbf{Building the Universal Mobility Model based on Global Data to Construct a Knowledge Base.} a) Developing a universal mobility framework, grounded in diverse data sources from around the globe, that is capable of generating OD flow information tailored to specific urban characteristics would enable widespread applicability across various cities~\cite{rong2024largescalebenchmarkdatasetcommuting}. b) Harnessing information from diverse sources, including traffic, ICT, demographics, census data, GIS, navigation systems, and mobile phone records, and amalgamating these disparate datasets into a unified framework would enhance overall data quality and complementarity. This approach would empower researchers to leverage the strengths of each data source, thereby generating synergistic effects that surpass the sum of their individual contributions. UrbanKG~(urban knowledge graph)~\cite{liu2023urbankg,wang2021spatio,liu2023knowledge,liu2021knowledge} may be a good choice for this kind of unified modeling. Once a model has been exposed to the global data, it can be regarded as a knowledge base that stores abundant information, which can be utilized for acquiring OD flow information in various scenarios.
\end{packed_itemize}

\subsubsection{Application-driven Edging Algorithm Introducing}

Certain practical application scenarios also drive researchers to choose more cutting-edge algorithms to address the corresponding challenges.
\begin{packed_itemize}
    \item \textbf{Utilizing Advanced Techniques to Address the Problem of Out-Of-Distribution When Transfer between Different Space.} In practical scenarios, data is often available in more developed regions, while data scarcity is a challenge in developing regions. There are significant pattern differences between the two, and utilizing causal learning~\cite{zeng2022causal} and transfer learning~\cite{rong2023goddag} to capture more fundamental correlations can address the issue of training models that are inconsistent with the distribution of the target scenarios.
    \item \textbf{Constructing Efficient Models to Deal With the Challenge of large scale.} Developing efficient models through simplification entails the formulation of elegant and streamlined models~\cite{rong2023complexityaware,bojchevski2018netgan} that strike a balance between complexity and performance while preserving computational efficiency and adhering to academic rigor. This will improve the scalability of methods, Incorporating hardware optimization and the design of high-performance computing systems, which will be essential approaches to address the challenges of high-dimensional OD matrix modeling.
\end{packed_itemize}

\subsubsection{Coupling With Urban Simulation}

Integrating OD flow data with computational models and urban simulation research is an important future direction. This direction primarily explores the relationship between data-driven computational models and urban simulation frameworks~\cite{yuan2024unist,yuan2024spatio}, empowering urban simulation with OD flows to obtain more realistic and reliable counterfactual decision-making references~\cite{zhang2024moss,zhang2024gpu,feng2024citygpt}. This emerging direction is currently underexplored, with a focus mainly on the application of reinforcement learning in traffic signal control issues~\cite{zheng2019learning}. Furthermore, with the OD flow data, a more fine-grained simulation of human behaviors in the urban context can be achieved, such as spatiotemporal event modeling~\cite{yuan2023spatiotemporal,yuan2022activity,feng2024citybench}, individual mobility modeling~\cite{long2023practical,wang2023pategail,yuan2023learning,xu2023urban}, pedestrian walking simulation~\cite{shi2023learning} and communication behaviors~\cite{hui2023large}, which will power up the urban simulation for a more generalized and reasonable deduction. Integrating the OD flow information in these modeling will guarantee the consistency of the urban behaviors from different scales.

\section{Conclusion and Summary} \label{sec:conclusion}

This survey provides a comprehensive overview of the relationships between OD flows and various disciplines, summarizing the topic from the perspectives of population mobility theory, techniques for specific problems, and practical applications based on OD flow information. In terms of theoretical research on mobility related to OD flows, we delved into the fields of urban geography, regional economics, and sociophysics, meticulously summarizing and organizing the research framework and development trajectory. We also identified four classic OD flow obtaining problems in real-world scenarios, along with a summary of their corresponding solutions and evaluation methods for different scenarios. Lastly, we discuss the support and benefit that OD flow information can offer to various disciplines in specific contexts and, based on existing challenges, outline future directions for development. We believe this survey will provide researchers from multiple disciplines with a comprehensive understanding of OD flows and their implications.

\begin{acks}
  This work is supported by a grant from the National Natural Science Foundation of China under 62272260 and U20B2060, the National Key Research and Development Program of China under 2020YFA0711403.
\end{acks}

\bibliographystyle{ACM-Reference-Format}
\bibliography{reference.bib}

\appendix

\section{Related Reviews} \label{apdx:related}

\begin{table*}[ht]
  \begin{center}
  \begin{tabular}{c|p{12cm}}
  \hline
   \textbf{Paper} & \textbf{Description} \\ \hline\hline
    \cite{luca2021survey}       & \tabincell{l}{This scholarly work presents a comprehensive overview of the utilization of deep \\ learning methodologies for resolving human mobility dilemmas in urban contexts. \\ The issues are significantly explicated and analyzed within a framework consisting \\ of four correlated dimensions: individual and collective perspective, as well as the \\ problems of prediction and generation thereof.}     \\ \hline
    \cite{barbosa2018human}       & \tabincell{l}{This article provides a overview of research on human mobility, examining the topic \\ from both individual and group perspectives. The discussion centers around physical \\ theoretical laws and modeling, with relevant applications in specific scenarios.}     \\ \hline\hline
    \cite{wilson1971family}       & \tabincell{l}{This review provides a comprehensive and in-depth review of the development of \\ the gravity model, which plays an important role in spatial interaction modeling.}     \\ \hline
    \cite{roy2003spatial}       & \tabincell{l}{This review provides an account of the evolution of spatial interaction modeling, \\ encompassing advancements in analogical models beyond the traditional gravity \\ -based principles, particularly the utilization of the more comprehensive concepts of \\ entropy and information theory.}     \\ \hline\hline
    % \cite{haynes2020gravity}       & \tabincell{l}{Gravity and Spatial Interaction Models.}     \\ \hline\hline
    \cite{puignau2020new}       & \tabincell{l}{This literature review provides an overview on contemporary data collection techni \\ -ques and advanced OD flow modeling, which can enhance our understanding of \\ population mobility patterns within cities.}     \\ \hline
    % \cite{}       & \tabincell{l}{Towards a generic benchmarking platform for origin–destination flows \\ estimation/updating algorithms: Design, demonstration and validation.}     \\ \hline
    \cite{hussain2021transit}       & \tabincell{l}{This article introduces the application of smart card data in estimating public transit \\ origin-destination matrices (tOD).}   \\ \hline
    \cite{bera2011estimation}       & \tabincell{l}{This survey provides an overview of the state-of-the-art technologies employed in \\ the different stages of estimating OD flows based on traffic counts on the roads.} \\ \hline
  \end{tabular}
  \end{center}
  \caption{Related reviews of research on origin-destination flows.}
  \label{Tab:existingsurvey}
\end{table*}

The two surveys \cite{luca2021survey,barbosa2018human} pertaining to human mobility delve into the study of the movement patterns of individuals and populations in cities. Luca et al.~\cite{luca2021survey} gave a review, whose primary focus revolves around the utilization of deep learning techniques to address various challenges and concerns pertaining to human mobility in urban settings. Barbosa et al.~\cite{barbosa2018human} offers a detailed examination of the physics-based theories and models that underpin various aspects of human mobility. Given the broad scope of human mobility, it is worth noting that the two works~\cite{luca2021survey,barbosa2018human} under discussion do not provide an extensive investigation of population mobility patterns between regions, i.e., OD flows. These works also devote substantial sections to the discussion of individual mobility as well as inflow and outflow. This represents a significant departure from the focus of this literature.

The existing literature on spatial interaction modeling~\cite{wilson1971family,roy2003spatial} is primarily focused on modeling the interaction between different regions of a city at a macro level, which is closely related to OD flows. However, these models are not limited to population movement, but also encompass a broader range of content such as goods, information, and other factors. These studies focus more on interpreting the intrinsic mechanisms of spatial interaction from a socio-economic perspective, in which gravity models~\cite{zipf1946p,roy2003spatial} play a significant role. This paper, on the other hand, focuses primarily on human mobility, with a more comprehensive exploration of theories, models, methods, and techniques than the aforementioned spatial interaction modeling reviews~\cite{wilson1971family,roy2003spatial}.

The final type of surveys~\cite{puignau2020new,hussain2021transit,bera2011estimation} primarily explore and summarize the methods and techniques used in the transportation field to obtain OD flows information. This has already been standardized as a common problem in transportation, known as OD estimation~\cite{van1980most,noursalehi2021dynamic,cascetta1993dynamic}. These issues are very specific, and in this paper, the development of technical solutions to address these problems and a systematic comparison will also be reviewed in this work.

\section{Data Details} \label{apdx:data}

\subsection{Mobility Data Related to OD Flows}

\begin{packed_itemize}
    \item \textbf{Travel Survey.} The travel survey~\cite{axhausen2002observing,schuessler2009processing} is a research methodology used to collect data on individuals' travel patterns and behavior. This type of survey typically involves the administration of questionnaires or interviews to a sample of individuals or households, and aims to gather information on various aspects of travel, such as trip purposes, modes of transportation used, trip distances, and travel times. Through the aggregation of all individual data collected, it is possible to calculate and derive important information such as daily OD flows or commuting OD flows, i.e., the number of home-work pairs.
    \item \textbf{Call Detail Records.} Call Detail Records (CDRs) typically contain detailed information about specific phone calls, including the time and date of the call, its duration, the caller's phone number, the recipient's phone number, and information about the base stations (or cell towers). By collecting and analyzing location information during phone calls from a large number of individuals over a prolonged period, it is possible to identify complete individual spatial patterns and movement characteristics~\cite{iqbal2014development,calabrese2011estimating}. 
    \item \textbf{Cellular Network Access.} Cellular Network Access data, like CDR data, records information about devices accessing cellular networks, including the spatial location of corresponding base stations. The biggest difference from CDR data is that the recording of location data is denser and the trajectory is much more continuous than that of CDR data~\cite{gundlegaard2016travel,pan2006cellular,yang2021detecting}.
    \item \textbf{GPS Records.} GPS data refers to location information collected by Global Positioning System (GPS) receivers~\cite{sadeghinasr2019estimating}. GPS is a satellite-based navigation system that provides precise geolocation and time information to GPS receivers anywhere on or near the Earth's surface. GPS data typically includes the latitude, longitude, and elevation of the receiver, as well as the time at which the data was collected~\cite{kaplan2017understanding}. GPS data possesses high accuracy and dense temporal and spatial resolution, which makes it valuable for various applications. However, due to privacy concerns and the considerable costs associated with data collection, storage, GPS data is often only available for a limited number of individuals, usually up to several thousand, which limits the scope and generalizability of research findings~\cite{kitchin2014real}.
    \item \textbf{Location-based Social Network Check-ins.} Location-based Social Network (LBSN) check-ins data refers to the records of users' location-specific activities published on social networking platforms such as Facebook, Foursquare, and Twitter. When users "check-in" at a particular location, they voluntarily provide information about their whereabouts and activities, and any accompanying text or media content. Check-ins data has the advantages of high location accuracy and large-scale coverage of user populations. However, because the data is uploaded voluntarily by users, there is often a sampling bias and the temporal and spatial granularity of the data can be sparse~\cite{noulas2012tale}.
    \item \textbf{Traffic Surveillance Video Data.} Traffic Surveillance Video Data refers to a collection of video recordings that capture the movement and behavior of vehicles and pedestrians on public roads, highways, and other transportation infrastructures. These video data are typically obtained from surveillance cameras installed at strategic locations, such as intersections, toll booths, and highways. Through such kind of vehicular movements in transportation infrastructure, it is possible to derive OD flow information for vehicles~\cite{goulart2017traffic}.
    \item \textbf{Smart Card.} Smart card data mainly refers to transaction data generated by using smart cards to pay fares in public transportation systems. Smart cards are integrated circuit cards that can store and process data, and are widely used in public transportation systems such as subways, buses, and light rails~\cite{pelletier2011smart}. It reflects passengers' demand for public transportation.
    \item \textbf{Taxi Orders.} Taxi order data refers to the data generated by the process of ordering and dispatching taxis through online or mobile platforms. The data usually contains information such as pick-up/drop-off location, pick-up/drop-off time, taxi type, payment method, and driver information. This data provides valuable insights into the spatial and temporal travel patterns of taxi users, as well as their travel preferences and behaviors. The analysis of taxi order data can help researchers and transportation planners better understand the demand for taxi services~\cite{wang2019origin}.
\end{packed_itemize}

\subsection{Auxiliary Data} \label{apdx:auxdata}

\begin{packed_itemize}
    \item \textbf{Demographics.} Demographics refer to statistical data that describe the characteristics of a population, such as age, gender, income, education, ethnicity, and occupation. The study of demographics is crucial in understanding the social and economic composition of a particular area or community. Individuals' mobility is contingent upon a range of factors, including age and income. For example, research suggests that younger and higher-income populations tend to exhibit greater levels of mobility compared to their older and lower-income counterparts~\cite{paez2012measuring}. 
    \item \textbf{Land use.} The distribution of land use types within a given region serves as an important indicator of its functional role within the urban landscape~\cite{anas1998urban}. Specifically, regions characterized by a higher percentage of residential land use are more likely to support nocturnal activities associated with sleep and rest, whereas those with a greater proportion of commercial land use are more likely to facilitate employment, shopping, and leisure activities. This relationship between land use and activity patterns also has implications for mobility, with a larger number of individuals typically returning to regions with a higher proportion of residential land use during nighttime hours. Consequently, the examination of land use patterns in urban environments is a crucial consideration in the analysis of mobility and its associated socio-economic and spatial dynamics.
    \item \textbf{POIs.} Points of interest (POIs) refer to locations that are of particular importance or interest within a given study geographical area. Examples of POIs include landmarks, public facilities, commercial establishments, and cultural institutions. POIs are typically identified based on their significance to the local community and their potential to attract visitors or customers. In urban contexts, POIs play an important role in shaping the activity patterns and mobility of the population.
    \item \textbf{Infrastructure.} Infrastructure refers to the basic physical and organizational structures and facilities that are necessary for the functioning of a society. In urban areas, infrastructure plays a critical role in shaping patterns of mobility and facilitating the provision of public services. What is more, transportation-related infrastructure is widely recognized as a key determinant of mobility.
\end{packed_itemize}

\begin{table}[]
\begin{tabular}{@{}clcc@{}}
\toprule
Collection      & \multicolumn{1}{c}{Information}     & \multicolumn{1}{c}{Objective} & \multicolumn{1}{c}{Level} \\ \midrule
Road Sensors    & \tabincell{l}{Traffic speed/counts} & \tabincell{l}{Vehicles}       & Group                     \\
RFID            & \tabincell{l}{Vehicle trajectories} & \tabincell{l}{Vehicles}       & Individual                \\
Station Sensors & \tabincell{l}{Visitor counts}       & \tabincell{l}{Passengers}     & Group                     \\
Smart Card      & \tabincell{l}{In/out check-ins}     & \tabincell{l}{Passengers}     & Individual                \\ \bottomrule
\end{tabular}
\caption{The comparison of traffic information collection methods.}
\label{tab:trafdata}
\end{table}

\section{Analytical Theories}

\subsection{Gravity Model with Different Constraints}

\subsubsection{Globally-constrained Gravity Model}

According to Equation 3, it can be obtained that:
\begin{equation}
    \hat{T}_{ij} = \lambda P_i^{\beta_i} P_j^{\beta_j} d_{ij}^{-\alpha}
\end{equation}
where $\lambda$, $\beta_i$, $\beta_j$ and $\alpha$ are all calibratable parameters the same as Eq. \ref{eq:genG}, and $\hat{T}_{ij}$ . Nevertheless, in this context, the parameter $\lambda$ assumes an additional function as a coefficient of proportionality that adjusts the overall magnitude of travel originated by the gravity model. It can be expressed as the following formula:
\begin{equation}
    \lambda = \frac{T}{\sum_{i=1}^m \sum_{j=1}^n P_i^{\beta_i} P_j^{\beta_j} d_{ij}^{-\alpha}}.
\end{equation}
In this equation, the parameters $\beta_i$, $\beta_j$ and $\alpha$ convey information about the spatial distribution of the studied phenomenon, which enables the determination of the relative magnitudes of OD flows between different regions. On the other hand, the parameter $\lambda$ provides information about the global total information, thereby determining the absolute magnitudes of all OD flows. 

\subsubsection{Production-constrained Gravity Model}

To use the outflow of each region as the constraints in the gravity model, we can combine it with Eq. \ref{eq:genG} to obtain:
\begin{equation}
    \hat{T}_{ij} = A_i O_i P_j^{\beta_j} d_{ij}^{-\alpha}
\end{equation}
where
\begin{equation}
    A_i = \frac{1}{ \sum_j P_j^{\beta_j} d_{ij}^{-\alpha} }
\end{equation}
So the final model can be expressed as the following equation.
\begin{equation}
    \hat{T}_{ij} = \frac{O_i P_j^{\beta_j} d_{ij}^{-\alpha}}{\sum_j P_j^{\beta_j} d_{ij}^{-\alpha}}.
\end{equation}

\subsubsection{Attraction-constrained Gravity Model}

Combining with Eq. \ref{eq:genG}, we can obtain the following expression.
\begin{equation}
    \hat{T}_{ij} = P_i^{\beta_i} B_j D_j d_{ij}^{-\alpha}
\end{equation}
where
\begin{equation}
    B_j = \frac{1}{\sum_i P_i^{\beta_i} d_{ij}^{-\alpha}}
\end{equation}
So the final format of the model is shown as follow.
\begin{equation}
    \hat{T}_{ij} = \frac{P_i^{\beta_i} D_j d_{ij}^{-\alpha}}{\sum_i P_i^{\beta_i} d_{ij}^{-\alpha}}.
\end{equation}

\subsubsection{Doubly-constrained Gravity Model}

The doubly-constrained gravity model can be termed as the following form:
\begin{equation} \label{eq:doubly-gravity}
    \hat{T}_{ij} = A_i O_i B_j D_j d_{ij}^{-\alpha},
\end{equation}
where
\begin{align}
    \begin{aligned}
        A_i &= \frac{1}{\sum_j B_j D_j d_{ij}^{-\alpha}}, \\
        B_j &= \frac{1}{\sum_i A_i O_i d_{ij}^{-\alpha}},
    \end{aligned}
\end{align}
where $A_i$ and $B_j$ are the balancing factors that control the model to satisfy the constraints. In practical applications, these two factors are commonly estimated via an iterative procedure, after which the model is employed to produce all OD flows.

\subsection{Theoretical Derivation Details}

\subsubsection{Zipf's Principle of Least Effort} \label{apdx:zipf}

Zipf systematically investigates the intrinsic dynamics of inter-regional interactions between urban areas by examining the economic trade perspective. First, Zipf posits that when individuals make behavioral decisions, they tend to gravitate towards the option that requires the least effort. To substantiate this point, Zipf utilized the power-law distribution of community sizes as evidence. Specifically, due to the Zipf's Principle of Least Effort, two forms of economic structures emerged in cities: 1) localizing economy (Force of Diversification), and 2) big city economy (Force of Unification). The first economy, the localizing economy, arises as people tend to live close to the source of raw materials, thereby minimizing transportation costs and the amount of work involved in the production process. The second economy, the big city economy, is formed as people seek to minimize transportation costs during the consumption process, gradually giving rise to large cities where goods produced can be delivered to consumers with the least transportation effort. These two economic paradigms lead to people congregating in either production sites or large cities, resulting in the formation of diverse communities. Due to the coexistence of the two economic paradigms and the mutual constraints between them, an equilateral hyperbola balance is established within the entire economic system, resulting in a population scale law characterized by the equation $r \cdot P = C$, where $r$ means the ordinal rank of the population of the community, $P$ denotes the population number of the community and $C$ is a constant. By taking the logarithm of both sides of the equation, we obtain $\log r + \log P = \log C$. Zipf demonstrated the reliability of this formula through various data sources~\cite{zipf1946p}. Thus, this demonstrated the validity of the Principle of Least Effort.

Based on Zipf's Principle of Least Effort, Zipf pointed out that people would naturally choose closer locations for interactions, thus deriving that spatial interactions are inversely proportional to the distance between two locations. In the relationship between spatial interaction and population distribution, Zipf utilized the ideal experiment of assuming that the population income and labor conditions in all regions are homogenized. Under this assumption, everyone would contribute equally to the intensity of spatial interaction, leading to the natural conclusion that the strength of spatial interaction is directly proportional to the population of both origin and destination. Although this assumption does not conform to reality, such an ideal experiment still helps deepen the understanding of the gravity model's patterns observed in OD flows.

Zipf only proposed the direct and inverse proportional relationship between population, distance, and spatial interaction strength, without delving into a precise mathematical model. By drawing an analogy to Newton's law of gravitation, Zipf opened the door to a series of research on gravity models. 

\subsubsection{Entropy Approach from Statistical Mechanics} \label{apdx:entropy}

In statistical mechanics, entropy is a fundamental concept that quantifies the degree of disorder or randomness in a system at the microscopic level. It serves as a measure of the number of possible microscopic states or configurations, known as microstates, which are consistent with the macroscopic properties of the system, such as energy, temperature, and pressure. The entropy of statistical mechanical introduced by Ludwig Boltzmann in the late 19th century, is given by the famous Boltzmann entropy formula:
\begin{equation}
    S = k_B \ln (\Omega),
\end{equation}
where $S$ means the entropy of the system, $k_B$ denotes the Boltzmann constant, and $\Omega$ is the number of microstates corresponding to a given macroscopic state. 

In the field of human mobility, from a macroscopic perspective, the same OD matrix may correspond to multiple different combinations of individual travel decision states. The number of such states is measured by entropy. It is evident that this notion of entropy is directly borrowed from the concept of entropy in statistical mechanics, to describe the degree of uncertainty associated with a macroscopic OD matrix state. In detail, within Wilson's framework, given the known observed information as constraints, we can enumerate all possible combinations of individual mobility decisions that form the OD matrix using the classical probabilistic method. By maximizing entropy, we can then select the most likely macroscopic state corresponding to the OD matrix.

Below, we provide a formalized description proposed by Wilson, using the doubly-constrained gravity model as an example. Within a given time window, we know the inflow and outflow quantities for each region, as shown in Eq. \ref{eq:doubly}. The objective is to find the most probable OD matrix, which denotes as $\mathbf{T}$. In other words, from the perspective of combinatorics, this means finding the OD matrix $\mathbf{T}$ with the greatest number of microstate combinations. This can be expressed using the following formula:
\begin{equation} \label{eq:entropy}
    \Omega (\mathbf{T}) = \frac{T!}{T_{11}!(T-T_{11})! } \frac{(T-T_{11})!}{T_{12}!(T-T_{11} - T_{12})!} ...= \frac{T}{\prod_{ij}T_{ij}!},
\end{equation}
where $\Omega (\mathbf{T})$ means the number of microscopic states of OD matrix $\mathbf{T}$. In elucidating the formula, ${T!} / {T_{11}!(T-T_{11})! }$ denotes the quantity of potential combinations arising from selecting $T_{11}$ individuals out of the total $T$ individuals, whereas ${(T-T_{11})!} / {T_{12}!(T-T_{11} - T_{12})!}$ signifies the quantity of potential combinations resulting from choosing $T_{12}$ individuals from the remaining $(T-T_{11})$ individuals, and so on. In each instance, the selections are independent events; thus, the aggregate number of possible combinations of the whole OD matrix $\mathbf{T}$ is the cumulative product of the number of combinations of these individual selections, as illustrated by Eq. \ref{eq:entropy}. Therefore, to maximize the entropy of a macroscopic state corresponding to an OD matrix $\mathbf{T}$, we simply need to maximize the following expression.
\begin{equation}
    S_\mathbf{T} = k \ln (\Omega (\mathbf{T})),
\end{equation}
where $k$ means a parameters, which does not affect the maximization of entropy. Incorporating the constraints expressed in Eq. \ref{eq:doubly}, we can get the following optimization problem.
\begin{equation}
    \begin{aligned}
        \text{maximize} \quad & S_\mathbf{T} \\
        \text{subject to} \quad & \sum_{j=1}^n T_{ij} = O_i  \quad for \quad i = 1, 2, ..., m  \\
                                & \sum_{i=1}^m T_{ij} = D_j  \quad for \quad j = 1, 2, ..., n  \\
    \end{aligned}
\end{equation}
As a result, we can obtain
\begin{equation}
    T_{ij} = A_i O_i B_j D_j d_{ij}^{-\alpha},
\end{equation}
where
\begin{align}
    \begin{aligned}
        A_i &= \frac{1}{\sum_j B_j D_j d_{ij}^{-\alpha}}, \\
        B_j &= \frac{1}{\sum_i A_i O_i d_{ij}^{-\alpha}}.
    \end{aligned}
\end{align}
We can infer from this observation that it is consistent with the formula of the doubly-constrained gravity model obtained through analogizing Newton's law of universal gravitation, which is shown as Eq. \ref{eq:doubly-gravity}.

In summary, the maximum entropy theory based on statistical mechanics can provide theoretical support for the gravity model. However, it should be noted that this theory also involves some unrealistic assumptions, such as assuming all microstates to be equiprobable. Furthermore, when additional prior information is available about the space being modeled, this method may not have proper ways to further enhance the precision of the model.

\subsubsection{Minimization of Information Gain from Information Theory} \label{apdx:InfoGain}

In information theory, information gain is used to describe the difference in information between posterior probabilities and prior probabilities, where the posterior probabilities are obtained by updating the prior probabilities with new collected data. The formula for information gain is as follows:
\begin{equation}
    I(p_{1},p_{0}) = \log \frac{p_{1}}{p_{0}},
\end{equation}
where $p_1$ denotes the posteriori probability of an event and $p_0$ means the priori probability.

The information contained in OD flows encapsulates two dimensions: 1) the spatial probability distribution pertaining to population mobility, and 2) the magnitude of the flows. By normalizing the OD matrix, it is feasible to derive the probability matrix representing population movement among distinct regions within the spatial domain. The specific calculation is as follows:
\begin{equation} \label{eq:infoGain}
    p_{ij} = \frac{T_{ij}}{T},
\end{equation}
where $p_{ij}$ denotes the probability of one individual's transition starting from origin $i$ and ending at destination $j$. If we use $\{q_{ij}\}$ to describe the probabilities between regions observed in the most recent period, then Eq. \ref{eq:infoGain} can be adapted to describe the information gain for the OD matrix, which is shown as below.
\begin{equation}
    I(P,Q) = \sum_{ij} p_{ij} \log (p_{ij} / q_{ij}),
\end{equation}
where $P$ denotes the probability distribution information contained in the OD matrix, and $Q$ denotes corresponding part of observations. By minimizing the information gain $I(P,Q)$, our model can be brought closer to the distribution of the observed data.

Therefore, solving the spatial distribution of population movements embedded in the OD matrix becomes an optimization problem, which can be formulated with the combination of constraints represented by Eq. \ref{eq:doubly}, as follows:
\begin{equation}
    \begin{aligned}
        \text{minimize} \quad & I(P,Q) \\
        \text{subject to} \quad & \sum_{j=1}^n T_{ij} = O_i  \quad for \quad i = 1, 2, ..., m  \\
                                & \sum_{i=1}^m T_{ij} = D_j  \quad for \quad j = 1, 2, ..., n  \\
    \end{aligned}
\end{equation}
Through transformation, the optimization problem can be reformulated as follows:
\begin{equation}
     p_{ij} = q_{ij} a_i o_i b_j d_j,
\end{equation}
where
\begin{equation}
    \begin{aligned}
        \quad o_i = O_i / T, \\
        \quad d_j = D_j / T
    \end{aligned}
\end{equation}
and the estimation for parameters $a_i$ and $b_j$ is iteratively calculated using the following formula:
\begin{equation}
    \begin{aligned}
        a_i &= \frac{1}{\sum_j q_{ij}b_j d_j}, \\
        b_j &= \frac{1}{\sum_i q_{ij}a_i o_i}.
    \end{aligned}
\end{equation}

Interestingly, the approach commonly referred to as the \textit{Fratar method} was not introduced by a researcher named Fratar, but rather by L. Törnqvist in 1965. Additionally, this method is known by other names, such as the Furness method or Biproportional method. In the extant literature~\cite{snickars1977minimum}, it has been consistently evidenced that this particular methodology boasts commendable performance attributes. Nevertheless, it is imperative to highlight that the method omits the contemplation of travel expenses inherent in conventional gravity models, thereby excluding the factor of spatial distance from its purview. 

\subsubsection{Economic Principles of Utility Maximization} \label{apdx:utility}

We will elaborate on the detailed derivation process and the ultimate conclusion of this theory. Assume that a study area is divided into n+1n+1 regions, where region i=0i=0 denotes the origin and regions j=1,2,...,nj=1,2,...,n denote the destinations. Within the economic system constituted by this area, each individual acquires utility by relocating to other regions and engaging in interactions with unit people or commodities located at those regions. This utility $u_{ij}$ is represented by the subsequent equation.
\begin{equation} \label{eq:uij}
    u_{ij} = f(T_{ij}),
\end{equation}
where $u_{ij}$ denotes the net utility of one individual at origin ii and interacting with destination $j$, $f(\cdot)$ denotes the utility function, and $T_{ij}$ means the number of trips, i.e., interaction intensity. The aggregate utility derived from an individual's mobility is the cumulative sum of the utilities acquired through interactions with all potential destinations, as represented by the subsequent equation.
\begin{equation} \label{eq:ui}
    u_i = \sum_{j=1}^{n} f(T_ij),
\end{equation}
where $u_i$ means the aggregated utility of trips originating from $i$ and interacting with each unit in all destinations. Nonetheless, it is apparent that a destination encompasses multiple interactable units. Consequently, Eq. \ref{eq:ui} is refined to accommodate the realistic scenario in which numerous units at a destination are available for interaction. Assuming that the utility provided by each unit within the destinations is equal, we can conclude that the total utility of each destination is proportional to its population. Therefore, Eq. \ref{eq:ui} is adjusted to the following form.
\begin{equation}
    u_i = a\sum_{j=1}^{n} p_j f(T_ij),
\end{equation}
where $a$ is a proportional constant and $p_j$ denotes the population of the pepole or commodities located at destination $j$.

Moreover, human mobility across regions is not unconstrained; it is influenced by limitations associated with temporal and monetary costs. Thus, within the context of limited resources and costs, individuals are required to optimize their cumulative utility by engaging in spatial interactions with other regions in a restrained manner.
Niedercorn has taken into account both time and monetary costs in his theory, providing separate expressions to capture each of these dimensions, which are as follows.
\begin{equation}
    \begin{aligned}
        m_i \geq r \sum_{j=1}^n d_{ij} T_{ij}, \\
        h_i \geq \frac{1}{s} \sum_{j=1}^n d_{ij} T_{ij},
    \end{aligned}
\end{equation}
where $m_i$ denotes the total cost in money, $r$ represents the cost in money for moving one unit distance, $h_i$ stands for the limitation in time cost, $s$ means the speed, and $d_{ij}$ denotes the distance from origin $i$ to destination $j$. Finally, we can obtain the total net utility of an individual from all interactions, which is shown as below.
\begin{equation} \label{eq:Ui}
    U_i = a \sum_{j=1}^n p_j f(T_{ij}) - \lambda ( r \sum_{j=1}^n d_{ij} T_{ij} -m_i ),
\end{equation}
where $U_i$ means the total net utility and $\lambda$ denotes the Lagrangian multiplier. Here we take the example of monetary cost. Therefore, by maximizing the utility $U_i$, we can obtain the optimal decision for the travels $\{Tij | j=1,2,...,n\}$ to each destination $j$ from the origin $i$. 

To explore the suitability of different utility function forms, Niedercorn tested two specific forms: the logarithmic form and the power form. The equations for these two forms are presented below.
\begin{equation} \label{eq:2uF}
    \begin{aligned}
        f(T_{ij}) &= \ln T_{ij}, \\
        f(T_{ij}) &= T_{ij} ^ b \quad\quad\quad (0<b<1).
    \end{aligned}
\end{equation}
By substituting the above two utility functions into Eq. \ref{eq:Ui} and maximizing the individual's total net utility $U_i$, we obtain the following two optimization problems.
\begin{equation}
    \begin{aligned}
        \text{maximize} \quad & a \sum_{j=1}^n p_j \ln T_{ij} - \lambda ( r \sum_{j=1}^n d_{ij} T_{ij} -m_i ). \\
        \text{maximize} \quad & a \sum_{j=1}^n p_j T_{ij}^b - \lambda ( r \sum_{j=1}^n d_{ij} T_{ij} -m_i ).
    \end{aligned}
\end{equation}
By solving the two optimization problems mentioned above, two kind of expressions for $T_{ij}$ can be obtained.

\begin{equation}
    \begin{aligned}
        {\arg\max}_{T_{ij}} U_i(T_{ij}) = \quad & a \sum_{j=1}^n p_j \ln T_{ij} - \lambda ( r \sum_{j=1}^n d_{ij} T_{ij} -m_i ). \\
        {\arg\max}_{T_{ij}} U_i(T_{ij}) = \quad & a \sum_{j=1}^n p_j T_{ij}^b - \lambda ( r \sum_{j=1}^n d_{ij} T_{ij} -m_i ).
    \end{aligned}
\end{equation}
By solving the two optimization problems mentioned above, the corresponding expressions for $T_{ij}$ can be obtained.
\begin{equation} \label{eq:tij}
    \begin{aligned}
        T_{ij} &= (\frac{m_i}{r}) (\frac{p_j}{\sum_{j=1}^n p_j}) (\frac{1}{d_{ij}}) \quad\quad &\text{for logarithmic form utility function.} \\
        T_{ij} &= (\frac{m_i}{r}) (\frac{p_j^{1/(1-b)}}{\sum_{j=1}^n \frac{p_j^{1/(1-b)}}{d_{ij}^{b/(1-b)}}}) (\frac{1}{d_{ij}^{1/(1-b)}}) &\text{for power form utility function.}
    \end{aligned}
\end{equation}
Assuming all individuals in each origin $i$ follow the same strategy show in Eq. \ref{eq:tij}, the resulting $T_{ij}$ for all people will be magnified by the order of the number of population in origin $i$, which can be expressed as follow.
\begin{equation} \label{eq:TIJ}
    \begin{aligned}
        T_{ij} &= (\frac{M_i}{r}) (\frac{p_j}{\sum_{j=1}^n p_j}) (\frac{1}{d_{ij}}) \quad\quad &\text{for logarithmic form utility function.} \\
        T_{ij} &= (\frac{M_i}{r}) (\frac{p_j^{1/(1-b)}}{\sum_{j=1}^n \frac{p_j^{1/(1-b)}}{d_{ij}^{b/(1-b)}}}) (\frac{1}{d_{ij}^{1/(1-b)}}) &\text{for power form utility function.}
    \end{aligned}
\end{equation}
where
\begin{equation}
    \frac{M_i}{m_i} = p_i,
\end{equation}
in which $p_i$ denotes the population of origin $i$ and $M_i$ means the total monetary cost of all people in origin $i$.

Based on the expressions of $T_{ij}$ describing the spatial distribution of population mobility under the two utility function forms presented by Eq. \ref{eq:TIJ}, it can be observed that the spatial interaction intensity between regions, as represented in Eq. \ref{eq:zipf}, is directly proportional to the population size of the regions and inversely proportional to the distance between them. Niedercorn's investigation~\cite{niedercorn1969economic} departs from earlier studies that examined the gravity models through physical or information-theoretic lenses. His approach is rooted in economics, postulating that individuals seek to optimize their utility from spatial interactions, which subsequently leads to the derivation of the gravity law.

\section{Evaluation of Obtained origin-destination Information} \label{apdx:metrics}

The evaluation methods for the OD flow-related models can be divided into two main categories. The first one is when the ground truth can be used for evaluation. In this case, the available data is typically divided into training and testing sets, with the models being trained on the training set and their performance being tested on the testing set. The second category is when ground truth cannot be obtained, such as in scenarios where the OD flow information is derived from some OD construction methods. In these cases, the reliability of the information can only be assessed indirectly through other information, such as traffic observations and census data.

\subsection{Evaluation with Ground Truth Based on Nummerical  Metrics}

\begin{table}[]
\resizebox{11cm}{!}{
\begin{tabular}{@{}llc@{}}
\toprule
Indicator       & Discription & Calculation \\ \midrule
RMSE   & \tabincell{l}{This metric measures the average squared differ \\ -ences between predicted and actual values.}            & $ \sqrt{{\frac{1}{|\mathbf{F}|}}{\sum\nolimits_{r_i,r_j\in{\mathcal{R}}}{||}{{{\mathbf{F}}_{r_i,r_j}}-{\hat{\mathbf{F}}_{r_i,r_j}}{{||}_2^2}}}}, $            \\
MAE    & \tabincell{l}{This metric represents the average absolute diff \\ -erences between predicted and actual values.}            & $ {\frac{1}{|\mathbf{F}|}} \sum\nolimits_{r_i,r_j\in{\mathcal{R}}} | {{\mathbf{F}}_{r_i,r_j}}-{\hat{\mathbf{F}}_{r_i,r_j}} | $            \\
sMAPE  & \tabincell{l}{This metric measures the percentage of the aver \\ -age absolute differences between predicted and \\ actual values relative to the average of the actual \\ and predicted values.}            & $ {\frac{100\%}{|\mathbf{F}|} \sum_{r_i,r_j} \frac{{\mathbf{F}_{r_i,r_j}} - {\hat{\mathbf{F}_{r_i,r_j}}}}{(|\mathbf{F}_{r_i,r_j}|+|\hat{\mathbf{F}_{r_i,r_j}}|)/2}} $   \\
CD     & \tabincell{l}{This metric quantifies the angular similarity bet \\ -ween predicted and actual values by considering \\ the cosine of the angle between their vectors.}            & $ 1 - \frac{\mathbf{F} \cdot \mathbf{\hat{F}}}{\lVert\mathbf{F}\rVert \lVert\mathbf{\hat{F}}\rVert}
 $            \\ \hline
CPC    & \tabincell{l}{This metric measures the degree of overlap bet \\ -ween predicted and actual commuting patterns \\ by evaluating the shared commuting links among \\ them.}            & $ \frac{2\!\! \sum_{r_i,r_j\in{\mathcal{R}}}    \min(\mathbf{F}_{r_i,r_j}, \hat{\mathbf{F}}_{r_i,r_j})}{({\sum_{r_i,r_j\in{\mathcal{R}}}{\!\mathbf{F}_{r_i,r_j}}+ \sum_{r_i,r_j\in{\mathcal{R}}}{\!\hat{\mathbf{F}}_{ij}} })} $            \\
PC     & \tabincell{l}{This metric measures the linear relationship bet \\ -ween predicted and actual values, with a higher \\ value indicating a stronger positive correlation.}            & $ \frac{\sum_{r_i,r_j\in{\mathcal{R}}} (\mathbf{F}_{r_i,r_j} - \bar{\mathbf{F}})(\mathbf{\hat{F}}_{r_i,r_j} - \bar{\mathbf{\hat{F}}})}{\sqrt{\sum_{r_i,r_j\in{\mathcal{R}}} (\mathbf{F}_{r_i,r_j} - \bar{\mathbf{F}})^2}\sqrt{\sum_{r_i,r_j\in{\mathcal{R}}} (\mathbf{\hat{F}}_{r_i,r_j} - \bar{\mathbf{\hat{F}}})^2}}
 $            \\
MSSIM  & \tabincell{l}{This metric takes the OD matrix as an image and \\ quantifies the similarity between two  images (the \\ generated OD matrix and ground truth) by consi \\ -dering structural information, luminance, etc.}            & $  \frac{1}{N} \sum_{i=1}^{N} \frac{(2\mu_{\mathbf{F}_i} \mu_{\mathbf{\hat{F}}_i} + C_1)(2\sigma_{{\mathbf{F}_i \mathbf{\hat{F}}_i}} + C_2)}{(\mu_{{\mathbf{F}_i}}^2 + \mu_{\mathbf{\hat{F}}_i}^2 + C_1)(\sigma_{\mathbf{F}_i}^2 + \sigma_{\mathbf{\hat{F}}_i}^2 + C_2)} $            \\
CosSim & \tabincell{l}{This is a measure of similarity between two non \\ -zero vectors, calculated by taking the cosine of \\ the angle between them.}            & $ \frac{\mathbf{F} \cdot \mathbf{\hat{F}}}{\lVert\mathbf{F}\rVert \lVert\mathbf{\hat{F}}\rVert} $            \\
JSD    & \tabincell{l}{This is a symmetric measure of the dissimilarity \\ between two probability distributions.}            & $ { \frac{\mathbf{KL}(P_{\mathbf{F}}||P_{\hat{\mathbf{F}}}) +  \mathbf{KL}(P_{\hat{\mathbf{F}}}||P_{\mathbf{F}})} {2}} $            \\
MMD    & \tabincell{l}{This is a statistical test for determining the simi \\ -larity between two probability distributions.}            & $ \lVert \mathbb{E}_{\mathbf{F} \sim P_{\mathbf{F}}}[k(\mathbf{F}, \cdot)] - \mathbb{E}_{{\mathbf{\hat{F}}} \sim P_{\mathbf{\hat{F}}}}[k({\mathbf{\hat{F}}}, \cdot)] \rVert_{\mathcal{H}_k}^2 $            \\ \bottomrule
\end{tabular}
}
\caption{A summary on the numerical evaluation metrics of OD flows.}
\label{tab:metrics}
\end{table}

In the process of quantitatively comparing the derived OD flows with the ground truth, there are predominantly two categories of metrics employed: Firstly, error metrics, which signify the magnitude of deviation between the predicted or estimated OD flows and the actual values - a larger discrepancy corresponds to a higher value. Secondly, similarity metrics are utilized to gauge the degree of resemblance between the obtained OD flows and the ground truth; a higher value implies increased similarity to the ground truth, subsequently leading to enhanced performance. The commonly used metrics are summarized in Table \ref{tab:metrics}. 

Root mean squared error~(RMSE) is a widely used metric for measuring the accuracy of a predictive model or estimator by calculating the square root of the average squared differences between the predicted and actual values. It is particularly useful in regression analysis, as it provides an aggregate measure of the overall error magnitude, with a larger RMSE indicating larger errors and a smaller RMSE indicating better model performance. One of the key advantages of RMSE is that it is expressed in the same units as the predicted and actual values, making it easier to interpret and compare with other metrics. However, it should be noted that RMSE is sensitive to outliers and may not always be the most appropriate metric for every situation.
Mean absolute error~(MAE) is a popular metric for measuring the accuracy of a predictive model or estimator by calculating the average of the absolute differences between the predicted and actual values, which is the same as RMSE. One advantage of MAE is that it is less sensitive to outliers than RMSE, making it a more robust metric in certain situations. However, MAE does not emphasize large errors as much as RMSE, so it may not be the best choice if large errors are particularly important to identify or penalize.
Symmetric mean absolute percentage error~(sMAPE) is a metric used to measure the accuracy of a predictive model or estimator by calculating the average of the absolute percentage differences between predicted and actual values, while treating under- and over-predictions symmetrically. Unlike MAE and RMSE, sMAPE expresses error as a percentage, making it easier to compare model performance across different scales or units. However, it should be noted that sMAPE can lead to misleading results when dealing with small or zero actual values, as the percentage error can become extremely large or undefined in these cases. In such situations, alternative metrics like MAE or RMSE might be more appropriate.
Cosine distance~(CD) is a metric used to measure the dissimilarity between two non-zero vectors by calculating one minus the cosine of the angle between them. Unlike Euclidean distance, cosine distance focuses on the orientation of the vectors rather than their magnitude. By normalizing the vectors before computing their cosine similarity, cosine distance effectively measures the angular distance between the vectors, which can be more informative and less sensitive to differences in magnitude.
Common part of commuting~(CPC) refers to a measure that quantifies the overlap in commuting patterns between different locations or areas, such as the number of individuals who share the same origin and destination for their daily commute. By analyzing the commonality in commuting patterns, transportation planners and researchers can gain insights into potential bottlenecks, areas requiring improved transportation infrastructure, and travel behavior trends. CPC can help identify similarities and differences in commuting flows, informing urban planning and transportation policies to better cater to the needs of the population.
Pearson correlation~(PC), also known as Pearson's $r$, is a statistical measure used to assess the strength and direction of a linear relationship between two continuous variables. Pearson correlation is sensitive to outliers and assumes a linear relationship between variables, which may not always be appropriate for every situation. In such cases, alternative similarity measures like cosine similarity may be more suitable.
Mean Structural SIMilarity~(MSSIM)~\cite{wang2004image} is a perceptual image quality assessment metric that quantifies the similarity between two images by considering structural information, luminance, and contrast, with the aim of mimicking the human visual system's perception of image quality. It provides a more comprehensive and perceptually relevant measure of image quality than traditional metrics like Mean Squared Error (MSE).
Cosine similarity~(CosSim) is a measure used to calculate the similarity between two non-zero vectors by taking the cosine of the angle between them. By normalizing the vectors before computing their cosine, the metric effectively measures the angular distance between the vectors, providing a more informative and less sensitive measure of similarity than Euclidean distance, especially in high-dimensional spaces. So it's suitable for evaluating the generated OD flows.
Jensen–Shannon divergence~(JSD) is a symmetric measure of dissimilarity between two probability distributions, often used in information theory, machine learning, and data analysis for comparing data. By being symmetric and always finite, JSD overcomes some of the limitations of KLD~(Kullback-Leibler divergence), making it particularly suitable for comparing probability distributions in various applications.
Maximum mean discrepency~(MMD) is defined as the distance between the means of the samples mapped into a Reproducing Kernel Hilbert Space (RKHS), which allows for the comparison of complex and non-linear structures in the data. MMD has several advantages, such as being a non-parametric test that does not rely on specific distribution assumptions and being sensitive to differences in both the shape and location of the distributions.

Apart from the above mentioned metrics, contemporary studies~\cite{saberi2018complex,barthelemy2011spatial,saberi2017complex} have advocated for the evaluation of OD flows derived from models by adopting a network-centric metrics. This is due to the recognition that scrutinizing OD matrices in the context of complex network can unveil crucial properties, such as the scale-free distribution of node degrees etc., which might not be discernible through traditional analytical methods. Saberi et al.~\cite{saberi2017complex} discusses the research on understanding human mobility patterns from a complex network perspective. By conducting comparative analyses between cities, the study uncovers statistical properties of travel networks. That work posits that urban transportation can be viewed as a complex, densely connected network, with each node representing a travel origin or destination. Edges between nodes carry weights, which are used to quantify the volume of travel on that particular route. Futhermore, Saberi et al.~\cite{saberi2018complex} employ various statistical measures, such as node degree, node flux, link weight, and betweenness centrality, as well as network features like Kullback-Leibler divergence and edge-based network dissimilarity between several OD prediction models and the ground truth, to conduct a comparative analysis. 

\subsection{Validation with Observational Profile Information of Mobility Based on Extending Modules}

In numerous instances, researchers are confronted with the challenge of evaluating the disparity between the OD flows obtained using specific methodologies and actual real-world conditions when access to ground truth data is unattainable. Nevertheless, several indirect or auxiliary strategies can be employed to assess the veracity of the extracted OD flows. Existing literature predominantly focuses on the following four key approaches:
\begin{packed_itemize}
    \item \textbf{Contrasting with small-sample travel surveys:} Although travel surveys may encompass only a minor portion of the population, their spatial distribution can, to a certain degree, reflect the overall population's OD flow distribution. If the spatial distribution of the OD flows derived through algorithms or models corresponds with that of the small-sample travel survey, it can serve as an indicator of the reliability of the extracted OD flow data. Studies employing this approach include but not limited: \cite{mamei2019evaluating,bachir2019inferring,tongsinoot2017exploring,pourmoradnasseri2019od,alexander2015origin,toole2015path,calabrese2011estimating,sadeghinasr2019estimating,fekih2021data,bonnel2015passive,bonnel2018origin,imai2021origin,cao2021day,wang2019extracting,yang2021detecting,yang2015origin,osorio2019social,ma2013deriving}.
    \item \textbf{Utilizing traffic simulators or traffic flow assignment models:} By incorporating the OD flows into the road network, investigators can examine whether the traffic conditions generated by the estimated OD flows exhibit consistency with those observed in real-life scenarios. A high degree of concordance implies that the extracted OD flows demonstrate a considerable level of credibility. Examples of studies employing this approach include but not limited: \cite{mamei2019evaluating,bachir2019inferring,gundlegaard2016travel,iqbal2014development,cao2021day,caceres2007deriving,wang2013estimating,hamedmoghadam2021automated,nigro2018exploiting,yang2017origin}.
    \item \textbf{Comparison with benchmark methods:} Evaluate the method's performance by comparing the results against those obtained using benchmark methods or models with established performance in similar contexts. If both sets of results exhibit a high degree of consistency, it suggests that the OD flow data obtained is reliable to a certain extent. Works that employ this scheme include but not limited: \cite{bachir2019inferring,gundlegaard2016travel,demissie2016inferring,heydari2023estimating,cao2021day,wang2019extracting}.
    \item \textbf{Downstream tasks:} There are also some studies that introduce downstream tasks to evaluate the authenticity of the extracted OD flow information, such as using OD flows for travel time estimation. If incorporating the extracted OD flow information leads to an improvement in the performance of the downstream task, it can indirectly attest to the reliability of the OD flow data. Works that use this approach include but not limited: \cite{iqbal2014development,cao2021day,wang2019extracting,yang2015origin,munizaga2012estimation,cerqueira2022inference,yang2018understanding,janzen2016estimating,nigro2018exploiting}.
\end{packed_itemize}

\end{document}